\documentclass[prc,aps,amsfonts,amsmath,arydshln,superscriptaddress,floatfix,twocolumn]{revtex4}
\usepackage{graphicx,float} 
\usepackage{rotating}
\usepackage{subfig}
\usepackage{braket}
\usepackage[colorlinks = true,
            linkcolor = blue,
            urlcolor  = blue,
            citecolor = blue,
            anchorcolor = blue]{hyperref}

\begin{document}

\title{Quadrupole and octupole collectivity and cluster structures in Neon isotopes}

\author{P. Marevi\'c}
\affiliation{CEA,DAM,DIF, F-91297 Arpajon, France}
\affiliation{Institut de Physique Nucl\'eaire, Universit\'e Paris-Sud, IN2P3-CNRS, \\
Universit\'e Paris-Saclay, F-91406 Orsay Cedex, France}
\author{J.-P. Ebran}
\affiliation{CEA,DAM,DIF, F-91297 Arpajon, France}
\author{E. Khan}
\affiliation{Institut de Physique Nucl\'eaire, Universit\'e Paris-Sud, IN2P3-CNRS, \\
Universit\'e Paris-Saclay, F-91406 Orsay Cedex, France}
\author{T. Nik\v{s}i\'c}
\affiliation{Department of Physics, Faculty of Science, University of Zagreb, \\
Bijeni\v{c}ka c. 32, 10000 Zagreb, Croatia}
\author{D. Vretenar}
\affiliation{Department of Physics, Faculty of Science, University of Zagreb, \\
Bijeni\v{c}ka c. 32, 10000 Zagreb, Croatia}

\begin{abstract} 
The lowest positive- and negative-parity bands of $^{20}$Ne and neutron-rich even-even Ne isotopes are investigated using a theoretical framework based on energy density functionals. Starting from a self-consistent relativistic Hartree-Bogoliubov calculation of axially-symmetric and reflection-asymmetric deformation energy surfaces, the collective symmetry-conserving states are built using projection techniques and the generator coordinate method. Overall a good agreement with the experimental excitation energies and transition rates is obtained. In particular, the model provides an accurate description of the excitation spectra and transition probabilities in $^{20}$Ne. The contribution of cluster configurations to the low-energy states is discussed, as well as the transitional character of the ground state. The analysis is extended to $^{22}$Ne and the shape-coexisting isotope $^{24}$Ne, and to the drip-line nuclei $^{32}$Ne and $^{34}$Ne. The role 
of valence neutrons in the formation of molecular-type bonds between clusters is discussed. 
\end{abstract}

\date{\today}

\maketitle

\section{Introduction}

The formation of cluster states, a transitional phenomenon between the quantum-liquid and solid phases in nucleonic matter, stellar matter and finite nuclei presents a very active topic experimental and theoretical research in nuclear physics and 
astrophysics \cite{beck10,beck12,beck14,horiuchi12,kimura16,tohsaki17,freer17,ebran17}. In addition to dedicated microscopic approaches that have been mainly applied to light nuclei, more recently clustering phenomena have also been analyzed using the universal framework of energy density functionals (EDFs) \cite{ebran17}. Very interesting results have been obtained but, although one can perform qualitative studies of the formation and evolution of cluster structures already on the mean-field level \cite{ebran12,ebran14a,ebran14b,girod13}, for a quantitative analysis that can be compared to experiment the basic EDF framework has to be extended by including collective correlations related to symmetry restoration and nuclear shape fluctuations. In the present study we develop an EDF-based framework that includes configuration mixing of angular momentum- and parity-projected axially-symmetric and reflection-asymmetric deformed mean-field states. The generator coordinate method (GCM) is employed in a systematic calculation of low-energy spectroscopic properties for the chain of Neon isotopes, starting from the self-conjugated $^{20}$Ne and extending to the drip-line nucleus $^{34}$Ne.  This analysis is entirely based on a universal EDF, without any parameter of the interaction, basis states or method adjusted specifically to nuclei under consideration.

The self-conjugate nucleus $^{20}$Ne exhibits admixtures of cluster configurations already in the ground state, that is, it is 
characterized by a transition between homogeneous nucleonic matter and cluster structures.  
Various theoretical approaches have been used to analyze the low-energy structure of $^{20}$Ne:
the angular momentum projected Hartree-Fock model~\cite{ohta04}, the resonating group method~\cite{matsuse75}, 
the $5\alpha$ generator coordinate method~\cite{nemoto75}, the antisymmetrized molecular dynamics (AMD) model~\cite{kanada95,kimura04,taniguchi04}, and the generalized Tohsaki-Horiuchi-Schuck-R\"opke (THSR) wave function model~\cite{zhou12}.
An interesting feature of this isotope is the dissolution of the reflection-asymmetric $\alpha + ^{16}$O structure in higher angular-momentum states by decreasing the equilibrium distance between two clusters, $\alpha$ and $^{16}$O. This is unexpected because centrifugal effects should in principle elongate the nucleus. Very recently a beyond mean-field study of reflection-asymmetric molecular structures and, in particular, of the anti-stretching mechanism in $^{20}$Ne has been performed based on the relativistic EDF framework \cite{zhou16}. It has been pointed out that a special deformation-dependent moment of inertia, governed by the underlying shell structure, could be responsible for the rotation-induced dissolution of the $\alpha + ^{16}$O cluster structure in the negative-parity states. Furthermore, the formation of the cluster structures in $N\neq Z$ nuclei includes, in addition to the $N=Z$ clusters, quasi-molecular bonding by the valence neutrons. 
One such example is the chain of even-even Ne isotopes that can be described as an 
$\alpha+^{16}O+xn$ system. The structure of the lightest isotope with such a structure, $^{22}$Ne, was previously  
analyzed with the AMD model~\cite{kimura07}, and both the molecular orbital bands and the $\alpha+^{18}$O molecular
bands were predicted.

This study  is organized as follows. In Sec. \ref{sec:theo} we briefly outline the theoretical framework of symmetry-conserving configuration mixing calculation based on nuclear EDFs. Section \ref{sec:results} presents an extensive analysis of the structure of low-energy positive- and negative-parity bands of $^{20-34}$Ne isotopes, and section \ref{sec:conclusion} summarizes the results. 
\section{Theoretical framework}
\label{sec:theo}
Nuclear energy density functionals (NEDFs) provide a global theoretical framework for studies of 
ground-state properties and collective excitations that is applicable across the 
entire nuclide chart, from relatively light systems to superheavy nuclei, and from the valley of 
$\beta$-stability to the nucleon drip-lines. Modern NEDFs are typically determined by about ten 
to twelve phenomenological parameters that are adjusted to a nuclear matter equation of state and
to bulk properties of finite nuclei. Based on this framework, various structure 
models have been developed that go beyond the mean-field approximation 
and include collective correlations related to restoration of broken symmetries and 
fluctuations of collective variables \cite{niksic11,egido16,bally14}. These models have become standard tools 
for nuclear structure calculations, providing accurate microscopic predictions for many 
low-energy nuclear phenomena.

The present study of cluster configurations in the Ne isotopic chain is based on the relativistic functional DD-PC1~\cite{niksic08}. 
Starting from microscopic nucleon self-energies in nuclear matter and empirical global properties of the nuclear matter equation of state, the coupling parameters of DD-PC1 were fine-tuned to the experimental masses of a set of 64 deformed nuclei in the mass regions 
$A\approx 150-180$ and $A\approx 230-250$. The DD-PC1 functional has been further tested in calculations of ground-state properties of medium-heavy and heavy nuclei, including binding energies, charge radii, deformation parameters, neutron skin thickness, and excitation energies of giant multipole resonances.  Furthermore,  a quantitative treatment of open-shell nuclei requires the inclusion of pairing correlations. The relativistic Hartree-Bogoliubov (RHB) framework \cite{VALR.05,Meng.06}, in particular, provides a unified description of particle-hole (ph) and particle-particle (pp) correlations on a mean-field level by combining two average potentials: the self-consistent mean field that encloses all the long-range ph-correlations, and a pairing field that sums up the pp-correlations. The ph effective interaction is derived from the DD-PC1 functional, while a pairing force separable in momentum space~\cite{duguet04,tian09}:
$\bra{k} V^{1^{S}_0} \ket{k^\prime} = -G p(k)p(k^\prime)$ is used in the pp channel. By assuming a simple Gaussian ansatz $p(k) = e^{-a^2 k^2}$, the two parameters $G$ and $a$ were adjusted to reproduce the density dependence of the gap at the Fermi surface in nuclear matter, as calculated with the Gogny D1S parameterization~\cite{Berger1991_CPC61-365}. The separable pairing force reproduces pairing properties in spherical and deformed nuclei calculated with the original Gogny D1S force, yet significantly reducing the computational cost.

The Dirac-Hartree-Bogoliubov equations are solved by expanding the nucleon spinors in the basis of an axially symmetric harmonic oscillator. The map of the energy surface as a function of quadrupole and octupole deformation is obtained by imposing constraints on the quadrupole $Q_{20}$ and octupole $Q_{30}$ moments. 
The method of quadratic constraint uses an unrestricted variation of the function 
\begin{equation}
\langle H \rangle + \sum_{\lambda = 2,3} C_{\lambda 0} \left( \langle \hat{Q}_{\lambda 0} \rangle - q_{\lambda 0} \right)^2,
\label{eq:quadrconstr}
\end{equation}
where $\langle H \rangle$ is total energy, $\langle \hat{Q}_{\lambda 0} \rangle$ denotes expectation values of the mass multipole operators $\hat{Q}_{\lambda 0} \equiv r^{\lambda} Y_{\lambda 0}$, $q_{\lambda 0}$ are the constrained values of multipole moments, and $C_{\lambda 0}$ the corresponding stiffness constants. In general, the values of the multipole moments $\langle \hat{Q}_{\lambda 0}\rangle$ coincide with the constrained values 
$q_{\lambda 0}$ only at the stationary point. The difference between a multipole moment
$\langle \hat{Q}_{\lambda 0}\rangle$ and the constrained $q_{\lambda 0}$ depends on the
stiffness constant. Smaller values of $C_{\lambda 0}$ lead to larger deviations of 
$\langle \hat{Q}_{\lambda 0}\rangle$ from the corresponding constrained values $q_{\lambda 0}$. Increasing the value of the stiffness constant, on the other hand, often destroys the convergence of the self-consistent procedure. This deficiency is resolved by implementing the augmented Lagrangian method \cite{Staszack2010_EPJA46-85}. In addition,  
the position of the center of mass coordinate is fixed at the origin to decouple the spurious states. In the following we will also use 
dimensionless deformation parameters $\beta_{\lambda}$ defined as:
\begin{equation}
\beta_{\lambda} = \frac{4 \pi}{3 A R^{\lambda}}q_{\lambda 0}, \quad R = r_0 A^{1/3}.
\label{eq:beta}
\end{equation}

To obtain quantitative predictions that can be compared to data, the self-consistent RHB approach has to be extended to include symmetry restoration and allow for nuclear shape fluctuations. This can be accomplished by configuration mixing of symmetry-conserving wave functions. Starting from a set of mean-field states $\ket{\phi(q)}$ that depend
on the collective coordinate $q$, one can build approximate eigenstates of the nuclear Hamiltonian. In the 
present study the basis states  $\ket{\phi(q)}$ are obtained by solving deformation-constrained RHB equations, that is,
the generator coordinate $q$ denotes the discretized deformation parameters $\beta_2$ and $\beta_3$.
Since the RHB states $\ket{\phi(q)}$ are not eigenstates of the angular momentum or parity operators, it is necessary to construct 
basis states with good angular momentum and parity that are used to diagonalize the nuclear Hamiltonian:
\begin{equation}
\ket{J M \pi ; \alpha} = \sum_j \sum_K f_{\alpha}^{JK \pi}(q_j) \hat{P}_{MK}^{J} \hat{P}^{\pi}\ket{\phi(q_j)}.
\label{eq:gcmansatz}
\end{equation}
$\hat{P}_{MK}^{J}$ denotes the angular momentum projection operator:
\begin{equation}
\hat{P}_{MK}^{J} = \frac{2J+1}{8\pi^2}\int \,d\Omega D_{MK}^{J*}(\Omega)\hat{R}(\Omega),
\label{eq:projection}
\end{equation}
where the integral is carried out over the three Euler angles $\Omega = (\alpha, \beta, \gamma)$, 
$D_{MK}^J(\Omega) = e^{-iM\alpha} d_{MK}^{J} (\beta) e^{-iK\gamma}$ is the Wigner's D-matrix \cite{varshalovich88}, and the active rotation operator reads $\hat{R}(\Omega) = e^{-i\alpha\hat{J}_z} e^{-i\beta\hat{J}_y} e^{-i\gamma\hat{J}_z}$. 
Good parity quantum number is restored by choosing the reflection-symmetric basis, that is, by ensuring that for each $(\beta_2, \beta_3)$ state the basis always contains the corresponding $(\beta_2, -\beta_3)$ state as well.
Taking into account axial symmetry imposed on the RHB basis states ($\hat{J}_z \ket{\phi(q_j)} =0, \forall j$),
the integral in Eq. (\ref{eq:projection}) simplifies considerably, since the integrals over the Euler angles $\alpha$ and $\gamma$ can be carried out analytically. This, in turn, restricts the angular momentum projection to $K=0$ and the states in Eq. (\ref{eq:gcmansatz}) from now on read $\ket{J \pi;\alpha}$.
Additionally, an approximate particle number correction is performed 
by applying the transformation of the Hamiltonian kernel introduced in Refs.~\cite{bonche90,rodriguez02}.

The weight functions  $f_{\alpha}^{J \pi}$ in Eq. (\ref{eq:gcmansatz}) are determined by the variational equation:
\begin{equation}
\delta E^{J \pi} = \delta \frac{\bra{J  \pi ; \alpha} \hat{H} \ket{J \pi ; \alpha} }{\braket{J  \pi ; \alpha | J  \pi ; \alpha}} = 0,
\end{equation}
that is, by requiring that the expectation value of the nuclear Hamiltonian in the state (\ref{eq:gcmansatz}) is stationary 
with respect to an arbitrary variation  $\delta f_{\alpha}^{J \pi}$.
This leads to the Hill-Wheeler-Griffin (HWG) equation \cite{hwg}:

\begin{equation}
\sum_j \left[ \mathcal{H}^{J \pi}(q_i,q_j) - E_{\alpha}^{J \pi} \mathcal{N}^{J  \pi}(q_i,q_j) \right] f_{\alpha}^{J \pi}(q_j) = 0.
\label{eq:hwg1}
\end{equation}
The norm kernel $\mathcal{N}^{J\pi}(q_i,q_j)$ and the Hamiltonian kernel $\mathcal{H}^{J \pi}(q_i,q_j)$
are given by the generic expression: 
\begin{align}
\nonumber
\mathcal{O}^{J\pi}(q_i,q_j) &= \frac{2J+1}{2} \delta_{M0} \delta_{K0}  \int_{0}^{\pi} \,d\beta \sin(\beta) d_{00}^{J*}(\beta) \\&  
\times\bra{\Phi(q_i)} \hat{O} e^{-i\beta\hat{J}_y} \hat{P}^{\pi} \ket{\Phi(q_j)}.
\label{eq:hamiltonian-normoverlap1}
\end{align}

\noindent The HWG equation (\ref{eq:hwg1}) presents a generalized eigenvalue problem, thus the functions $f_{\alpha}^{J \pi}(q_j)$ are not orthogonal and cannot be interpreted as collective wave functions for the variable $q$. Therefore, one rewrites Eq. (\ref{eq:hwg1}) in terms of another set of functions,
$g_{\alpha}^{J \pi}(q_j)$, defined by
\begin{equation}
g_\alpha^{J\pi}(q_i) = \sum_j{{(\mathcal{N}^{J\pi})}^{1/2}(q_i,q_j)f_{\alpha}^{J \pi}(q_j) }.
\end{equation}
The HWG equation now defines an ordinary eigenvalue problem:
\begin{equation}
\sum_j \tilde{\mathcal{H}}^{J \pi}(q_i, q_j) g_{\alpha}^{J \pi} (q_j) = E_{\alpha}^{J \pi} g_{\alpha}^{J \pi} (q_i),
\label{eq:hwg2}
\end{equation}
with 
\begin{align}
\nonumber
\tilde{\mathcal{H}}^{J \pi}(q_i, q_j) =& \sum_{k,l} \Big[ (\mathcal{N}^{J \pi})^{-1/2}(q_i, q_k) 
\mathcal{H}^{J \pi}(q_k, q_l) \\& \times (\mathcal{N}^{J \pi})^{-1/2}(q_l, q_j) \Big].
\label{eq:hwhammod}
\end{align}
The functions $g_{\alpha}^{J \pi}(q_j)$ are orthonormal and play the role of collective wave functions.  In practice, one first diagonalizes the norm overlap kernel:
\begin{equation}
\sum_{j} \mathcal{N}^{J \pi}(q_i,q_j) u_k^{J \pi}(q_j) = n_k^{J \pi} u_k^{J \pi}(q_i).
\label{eq:normdiag}
\end{equation}
Because the basis functions $\ket{\phi(q_i)}$ are not linearly independent, many of the norm overlap kernel
eigenvalues $n_k$ have values close to zero. The corresponding eigenfunctions $u_k(q_i)$ are rapidly oscillating and do not carry any physical information. However, such states can lead to numerical problems and thus need to be removed from the basis. The collective Hamiltonian is built from the remaining states
\begin{equation}
\mathcal{H}_{kl}^{J \pi c} = \frac{1}{\sqrt{n_k}} \frac{1}{\sqrt{n_l}} \sum_{i, j} u_k^{J \pi}(q_i) 
\tilde{\mathcal{H}}^{J \pi} (q_i, q_j) u_l^{J \pi}(q_j),
\label{eq:collham}
\end{equation}
and subsequently diagonalized
\begin{equation}
\sum_l \mathcal{H}_{kl}^{J \pi c} g_l^{J \pi \alpha} = E_{\alpha}^{J \pi} g_k^{J \pi \alpha}.
\label{eq:hwg3}
\end{equation}
The solution determines both the ground-state and the energies of the excited states, for each value of the
angular momentum $J$ and parity $\pi$. The collective wave functions $g_{\alpha}^{J \pi}(q_j)$ and 
weight functions $f_{\alpha}^{J \pi}(q_j)$ are calculated from the norm overlap eigenfunctions
\begin{equation}
g_{\alpha}^{J \pi} (q_i) = \sum_k g_k^{J \pi \alpha} u_k^{J \pi} (q_i),
\label{eq:gfun}
\end{equation}
and 
\begin{equation}
f_{\alpha}^{J \pi} (q_i) = \sum_k \frac{g_k^{J \pi \alpha}}{\sqrt{n_k^{J \pi}}} u_k^{J \pi} (q_i).
\label{eq:ffun}
\end{equation}
Once the weight functions $f_{\alpha}^{J \pi}(q_j)$ are known, it is straightforward to calculate all physical observables, e.g. transition probabilities and spectroscopic quadrupole moments~\cite{rodriguez02}.
The spectroscopic quadrupole moment of a state $\ket{J \pi; \alpha}$ is defined as:
\begin{align}
\nonumber
Q^{spec}_2(J&\pi, \alpha) = e \sqrt{\frac{16\pi}{5}} \begin{pmatrix}
J  & 2 & J \\
J & 0 & -J 
\end{pmatrix} \\ & \times \sum_{q_i, q_j} f_{\alpha}^{J \pi *}(q_i) \braket{J \pi q_i || \hat{Q}_2 || J \pi q_j} 
f_{\alpha}^{J \pi}(q_j).
\label{eq:qspec}
\end{align}
The reduced electric multipole transition probability for a transition between an initial state $\ket{J_i \pi_i; \alpha_i}$ and a final state $\ket{J_f \pi_f; \alpha_f}$ reads:
\begin{align}
\nonumber
B&(E\lambda; J_i \pi_i \alpha_i \rightarrow J_f \pi_f \alpha_f) = \frac{e^2}{2J_i+1} \\ & \times \left| \sum_{q_i, q_f}
f^{J_f \pi_f *}_{\alpha_f}(q_f) \langle J_f \pi_f q_f || \hat{Q}_{\lambda} || J_i \pi_i q_i \rangle f^{J_i \pi_i}_{\alpha_i}
(q_i) \right|^2.
\label{eq:be2}
\end{align}
We emphasize that, since these quantities are calculated in the full configuration space, there is no need to introduce effective charges and $e$ denotes the bare value of the proton charge. 

\begin{figure*}[!ht]
{\includegraphics[scale=0.31]{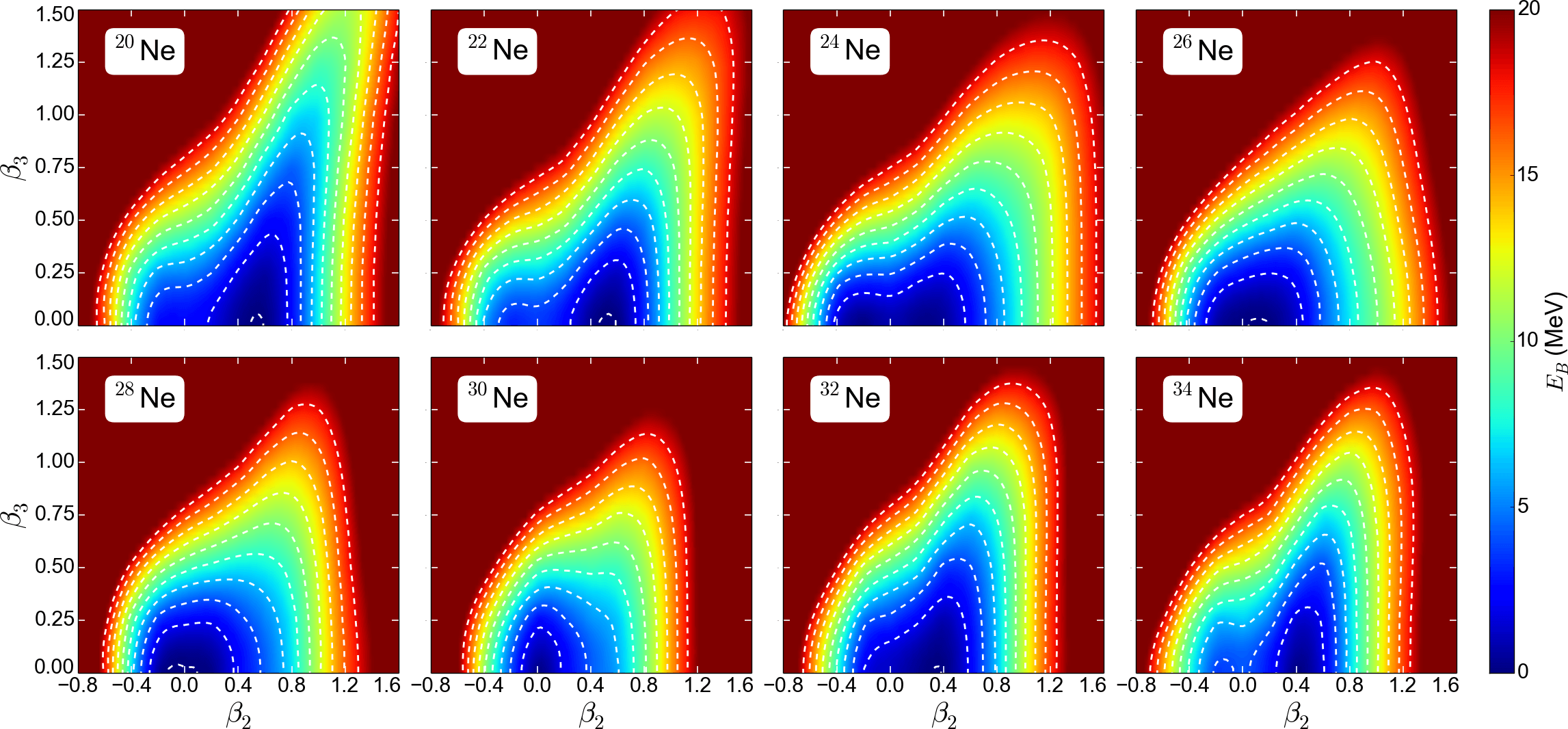}}
  \caption{(Color online) Self-consistent RHB binding energies of even-even $^{20-34}$Ne isotopes in the $\beta_2 - \beta_3$ plane. For each nucleus energies are normalized with respect to the corresponding absolute minimum. Dashed contours are separated by~$2$ MeV.}
    \label{fig:NeonPES}
\end{figure*}

\section{Evolution of cluster configurations in the Neon isotopic chain}
\label{sec:results}
\subsection{Spectroscopic properties of $^{20-34}$Ne}

Our analysis of the evolution of cluster configurations in the chain of isotopes $^{20-34}$Ne starts with a microscopic self-consistent relativistic Hartree-Bogoliubov calculation of quadrupole-octupole deformation energy surfaces. The Dirac-Hartree-Bogoliubov equations are solved by expanding nucleon spinors in the basis of an axially symmetric harmonic oscillator in cylindrical coordinates. To avoid the occurrence of spurious states, the large and small components of nucleon spinors are expanded in bases of $N_{sh}=10$ and $N_{sh}=11$ major oscillator shells, respectively~\cite{gambhir90}. The map of the energy surface as a function of quadrupole and octupole deformation is obtained by imposing constraints on the mass multipole moments $q_{20}$ and $q_{30}$ (cf. Eq. (\ref{eq:quadrconstr})). 

Figure \ref{fig:NeonPES} displays the RHB energy maps of the even-even  $^{20-34}$Ne isotopes in the $\beta_2 - \beta_3$ plane. For each isotope energies are normalized with respect to the absolute minimum. At the mean-field level the equilibrium state of all considered isotopes is axially symmetric. $^{20}$Ne and $^{22}$Ne exhibit prolate equilibrium minima with deformation $\beta_2 \approx 0.5$. By adding two more neutrons, an oblate-deformed minimum ($\beta_2=-0.27$) develops in $^{24}$Ne with an additional prolate-deformed local minimum ($\beta_2=0.28$). These two minima are separated in energy by only $240$ keV. Additional neutrons at first lead to nearly-spherical minima in $^{26}$Ne and $^{28}$Ne, and finally to a spherical equilibrium in $^{30}$Ne isotope caused by the $N=20$ neutron shell closure. Moving further away from the $N=20$ magic number, the neutron-rich isotopes $^{32}$Ne and $^{34}$Ne display prolate minima with deformation $\beta_2 = 0.33$ and $\beta_2 = 0.44$, respectively. It is interesting to note that the RHB model predicts for both of these isotopes to be stable against
the two-neutron emission, in agreement with data. The stability of Ne isotopes against neutron emission will be further analyzed using the configuration mixing framework.

In the next step, part of the symmetries broken on the mean-field level is restored by performing angular momentum and parity projection. The integrals involved in angular momentum projection are evaluated using an equidistant mesh for the Euler angle $\beta \in [0,\pi]$. We have verified that $N_\beta = 27$ mesh points ensures convergent results for all values of angular momenta $J\le 7$ and a broad range of quadrupole and octupole deformations.

 \begin{figure*}[] 
 {\includegraphics[scale=0.31]{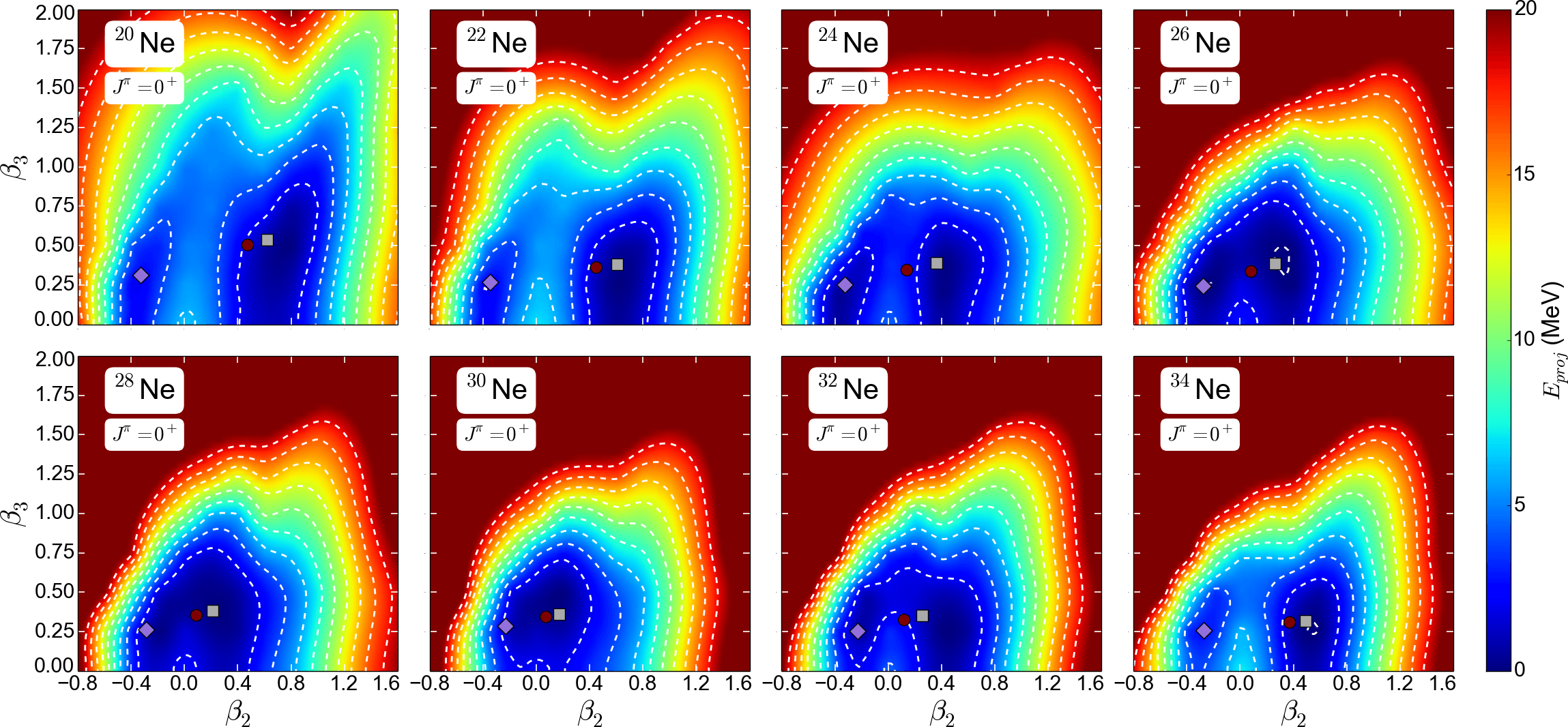}}
 \caption{(Color online) Angular momentum- and parity-projected energy surfaces of even-even $^{20-34}$Ne isotopes, for spin and parity $J^{\pi}=0^+$  in the $\beta_2 - \beta_3$ plane. For each nucleus energies are normalized with respect to the binding energy of the corresponding absolute minimum. Dashed contours are separated by $2$ MeV. See text for the explanation of symbols.}    
    \label{fig:projen0}
\end{figure*}

\begin{figure*}[]  
{\includegraphics[scale=0.31]{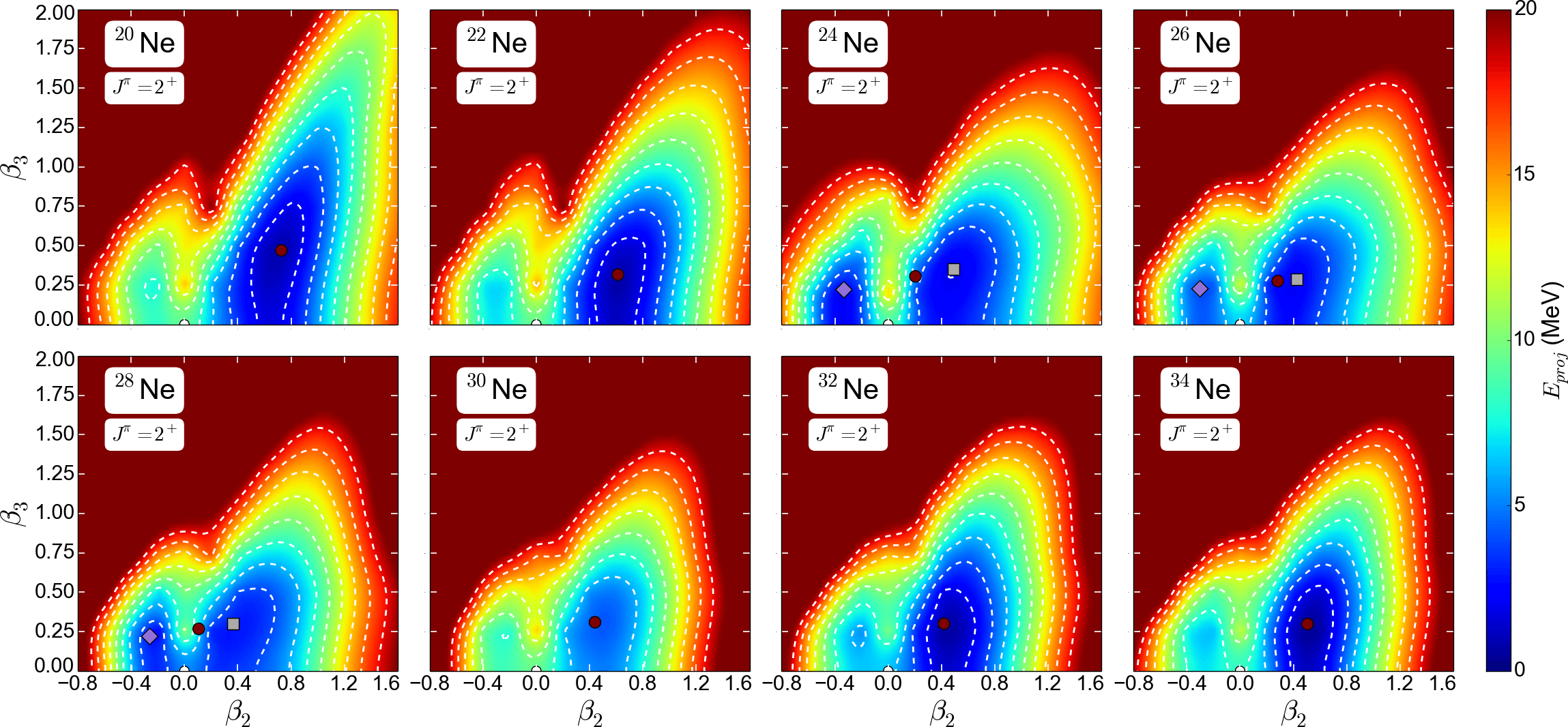}}
    \caption{(Color online) Same as in the caption to Fig.~\ref{fig:projen0}, but for spin and parity $J^{\pi}=2^+$.}    
    \label{fig:projen2}
\end{figure*}

Figures \ref{fig:projen0} and \ref{fig:projen2} show the angular momentum- and parity-projected energy maps for the  
positive parity states $J^{\pi}=0^+$ and $2^+$. For each isotope the energies are normalized with respect to the binding energy of the $J^\pi = 0^+$ minimum. We note that the angular momentum projection for the spherical $(\beta_2 = 0, \beta_3 = 0)$ configuration is well-defined only for $J^{\pi} = 0^+$, in other cases this point is omitted from the plots. In addition, on each energy map we have denoted by a circle the position of the average deformation for the lowest collective state obtained in the configuration mixing calculation. For collective states with significant contribution from oblate deformations ($\ge 10\%$), the positions of the average prolate and oblate deformations are denoted separately by the square and diamond symbols. A prominent feature in Fig.~\ref{fig:projen0} is the fact that parity projection shifts the position of the minimum towards octupole deformations. Angular momentum projection also modifies the topography of mean-field energy maps by lowering deformed configurations, thus forming additional local minima for all isotopes. For higher values of angular momentum ($J^{\pi}=2^+$, $J^{\pi}=4^+$, etc.), the absolute minimum is always prolate-deformed, except for the $^{24}$Ne and $^{28}$Ne isotopes that display shallow oblate minima for $J^{\pi} = 2^+$.

\begin{figure*}[]  
{\includegraphics[scale=0.31]{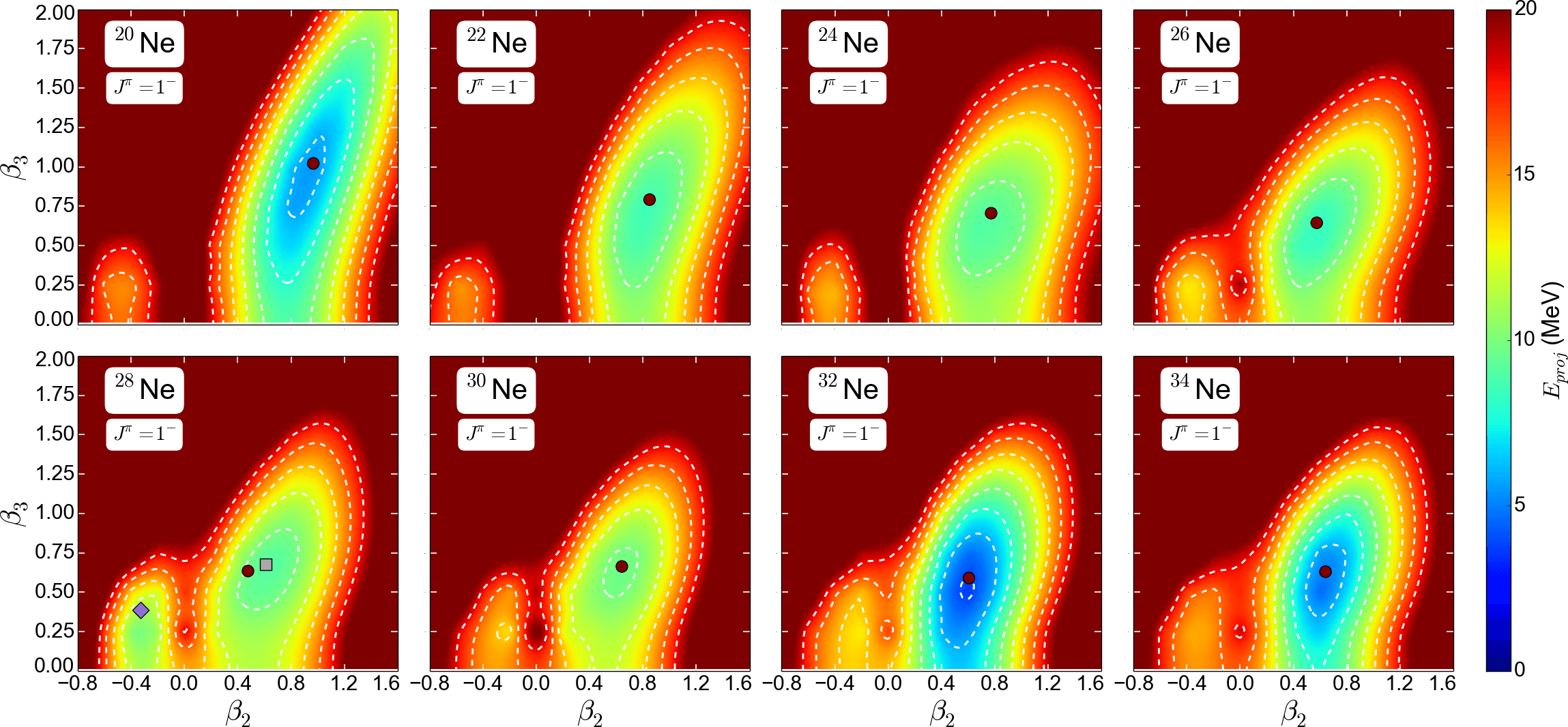}}
    \caption{(Color online) Angular momentum- and parity-projected energy surfaces of even-even $^{20-34}$Ne isotopes, for spin and parity $J^{\pi}=1^-$ in the $\beta_2 - \beta_3$ plane. For each nucleus energies are normalized with respect to the binding energy of the corresponding $0^+$ minimum. Dashed contours are separated by $2$ MeV. See text for the explanation of symbols.}    
    \label{fig:projen1}
\end{figure*}

\begin{figure*}[]  
 {\includegraphics[scale=0.31]{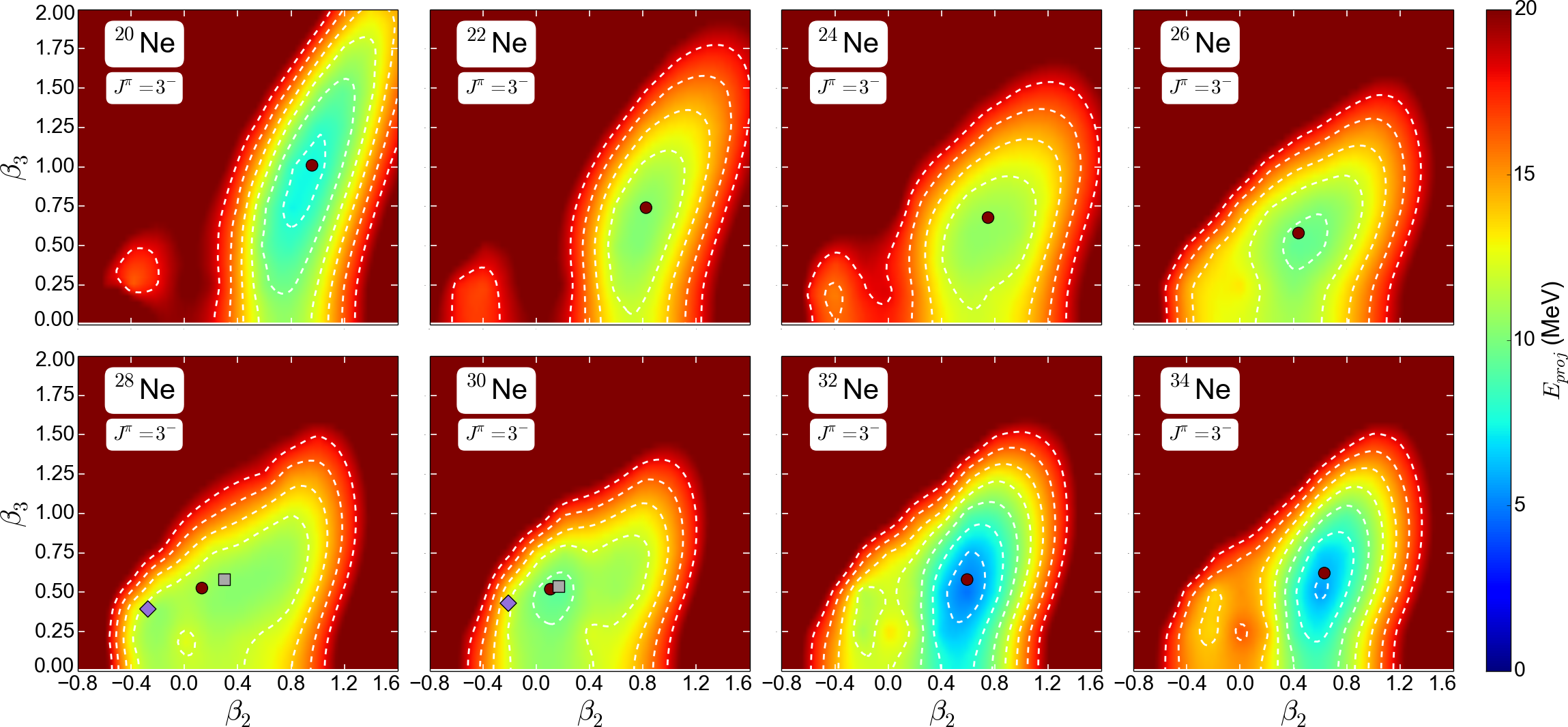}}
    \caption{(Color online) Same as in the caption to Fig.~\ref{fig:projen1}, but for spin and parity $J^{\pi}=3^-$.}    
    \label{fig:projen3}
\end{figure*}
 
In Figs.~\ref{fig:projen1} and \ref{fig:projen3} we plot the angular momentum- and parity-projected energy maps for the negative parity states $J^{\pi} = 1^-$ and $J^{\pi} = 3^-$. Again, for each isotope energies are normalized with respect to the binding energy of the minimum of the surface 
$J^\pi = 0^+$. Notice that parity projection for reflection-symmetric ($\beta_3 = 0$) configurations is well-defined only for positive parity, hence these configurations are omitted in Figs.~\ref{fig:projen1} and \ref{fig:projen3}. The negative parity-projected surfaces are rather soft in the octupole direction, with absolute minima located on the prolate side and separated at least by $4$ MeV from the $J^\pi=0^+$ minima. We note that angular momentum and parity projection modifies the topography of the energy maps and, therefore, indicates that configuration mixing calculations will play a crucial role for a quantitative description of the structure of Ne isotopes.

Correlation effects related to fluctuations of collective coordinates are taken into account by performing configuration mixing calculations of projected RHB states. The equidistant two-dimensional mesh covers a wide range of deformations in both the quadrupole and the octupole directions: $\beta_2 \in [-0.8,1.6]$ and  $\beta_3 \in [-2.0, 2.0]$. For the step size on the quadrupole and octupole grids we use 
$\Delta \beta_2 = 0.2$ and $\Delta \beta_3 = 0.25$, respectively. In addition, a cut-off in the RHB binding energy is introduced, that is, configurations with energy more than $30$ MeV above the RHB equilibrium state are not included in the GCM calculation. We have verified that this choice for the energy cut-off does not influence the final results. As a result, the number of basis states included in the configuration mixing calculation reads: $157$ ($^{20}$Ne), $149$ ($^{22}$Ne), $151$ ($^{24}$Ne), $143$ ($^{26}$Ne), $139$ ($^{28}$Ne), $123$ ($^{30}$Ne), $139$ ($^{32}$Ne) and $135$ ($^{34}$Ne).

Following the diagonalization of the norm overlap kernel, those eigenvectors which correspond to eigenvalues smaller than a given positive constant $\zeta$ are eliminated from the basis. This is necessary to prevent possible numerical instabilities occurring in the diagonalization of the collective Hamiltonian (see, for example, section 3.2. of Ref.~\cite{bonche90}). 
In Tab.~\ref{tab:zetaconv1} we show the calculated
ground-state energies for the even-even $^{20-34}$Ne isotopes as a function of the parameter $\zeta$. Obviously, for the values $\zeta = 5 \times 10^{-4}$ and $\zeta = 1 \times 10^{-3}$ the results are not stable, and the corresponding eigenvectors contain a considerable number of spurious components. For values between $\zeta = 5 \times 10^{-3}$ and $\zeta = 5 \times 10^{-2}$ stable results for the ground-state energy are obtained. Therefore, in all further calculations the cut-off parameter $\zeta = 5 \times 10^{-3}$ is used.

\begin{figure}[]  
      {\includegraphics[width=0.48\textwidth]{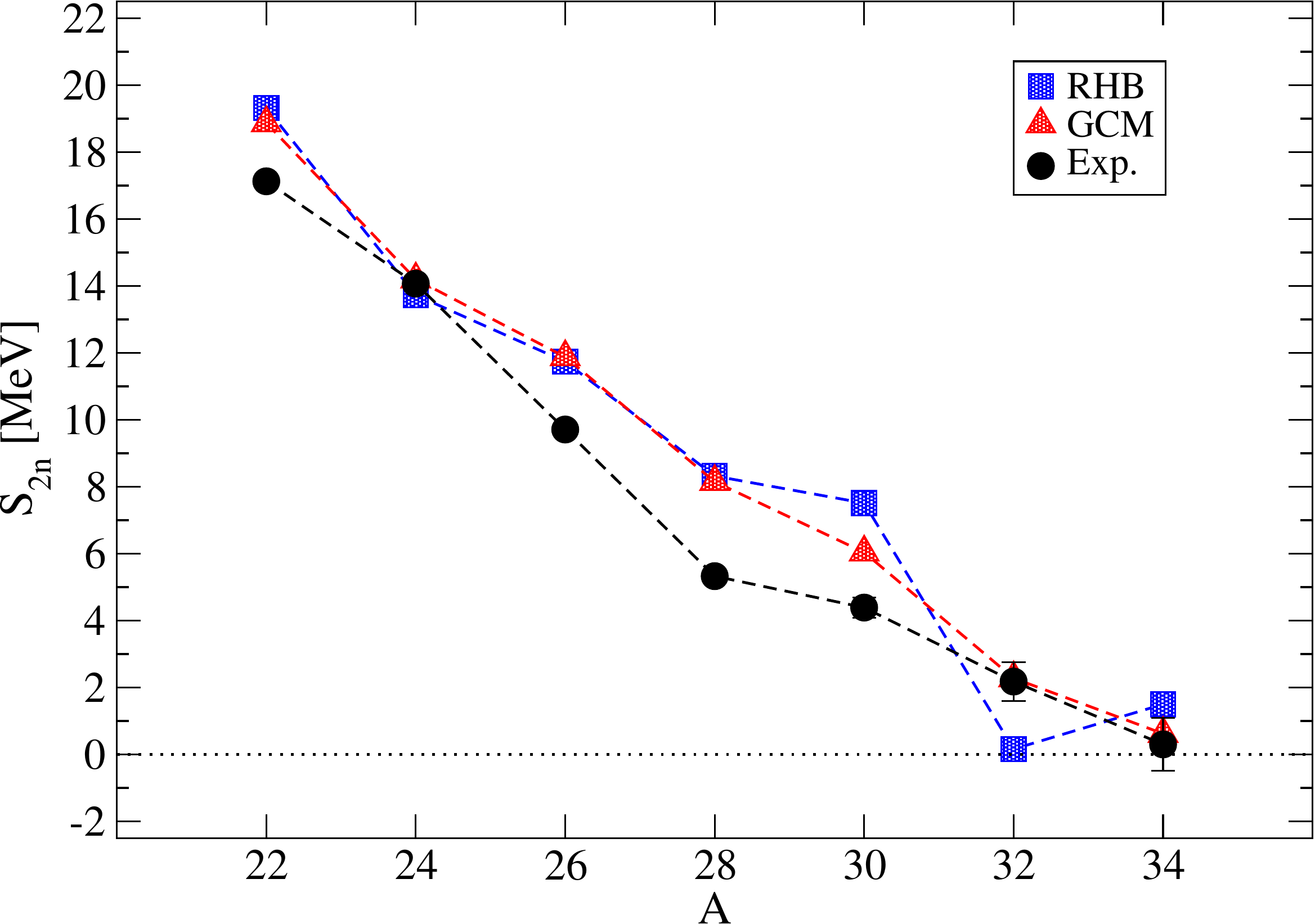}
    }
    \caption{(Color online) Two-neutron separation energies of $^{22-34}$Ne isotopes. The RHB values obtained on the mean-field level (squares), and results of the full angular momentum- and parity-projected GCM calculation (triangles), are compared to the available data \cite{wang12}.}    
    \label{fig:senergy}
\end{figure}

To analyze the predicted stability of Ne isotopes against two-neutron emission,  in Fig.~\ref{fig:senergy} we 
plot the two-neutron separation energies $S_{2n} = E_{0_1^+}(A-2,Z) - E_{0_1^+}(A,Z)$ for the even-even $^{22-34}$Ne isotopes. The full GCM configuration mixing results are compared with the available data~\cite{wang12} and, to quantify correlation effects, with the mean-field RHB results. The RHB results for the two-neutron separation energy, that is, the differences between binding energies of the corresponding equilibrium minima, generally overestimate the experimental values, except for $^{32}$Ne. It appears that for $A\le 30$ configuration mixing does not produce a significant impact on the calculated two-neutron separation energies. 
Closer to the drip-line, however, one notices that the inclusion of collective correlations through GCM configuration mixing becomes much more important and brings the theoretical $S_{2n}$ values within the experimental error bars. In addition, we have verified that $^{34}$Ne is the last Ne isotope predicted to be stable, since both the 
two-neutron ($S_{2n} = - 1.16$ MeV) and the 
four-neutron ($S_{4n} = -0.58$ MeV) separation energies for $^{36}$Ne isotope are negative. A similar improvement of the predicted two-neutron separation energies for Ne isotopes was also obtained in the angular momentum-projected GCM study of Ref.~\cite{rodriguez03}, based on the Gogny D1S effective interaction. However, the calculated $S_{2n}$ value for $^{34}$Ne in Ref.~\cite{rodriguez03} was slightly negative, that is, this isotope was predicted unstable against the two-neutron emission. 

\begin{table}[t]
\setlength{\tabcolsep}{4pt}
\renewcommand{\arraystretch}{1.1}
\caption{Calculated ground-state energies (in MeV) for the even-even $^{20-34}$Ne isotopes, as a function of the cut-off parameter $\zeta$ 
for the smallest eigenvalue of the norm overlap kernel matrix.}
\centering
\begin{tabular}{cccccc}
\hline \hline
$\zeta$ & $ 5 \times 10^{-4}$ & $ 1 \times 10^{-3}$ & $ 5 \times 10^{-3}$ & $ 1 \times 10^{-2}$ & $ 5 \times 10^{-2}$  \\ \hline
$^{20}$Ne & $-173.49$ & $-166.15$ &  $-162.49$ & $-162.46$  & $-162.35$ \\
$^{22}$Ne & $-202.69$ & $-192.70$ &  $-181.36$ & $-181.33$  & $-181.28$ \\
$^{24}$Ne & $-195.75$ & $-195.69$ &  $-195.56$ & $-195.51$  & $-195.47$ \\
$^{26}$Ne & $-207.59$ & $-207.54$ &  $-207.44$ & $-207.41$  & $-207.37$ \\
$^{28}$Ne & $-215.69$ & $-215.66$ &  $-215.59$ & $-215.56$  & $-215.39$ \\
$^{30}$Ne & $-221.82$ & $-221.76$ &  $-221.63$ & $-221.59$  & $-221.27$ \\
$^{32}$Ne & $-223.93$ & $-223.88$ &  $-223.81$ & $-223.72$  & $-223.55$ \\
$^{34}$Ne & $-224.46$ & $-224.46$ &  $-224.43$ & $-224.41$  & $-224.34$ \\ \hline \hline
\end{tabular}
\label{tab:zetaconv1}
\end{table}

\begin{figure*}[]  
      {\includegraphics[scale=0.32]{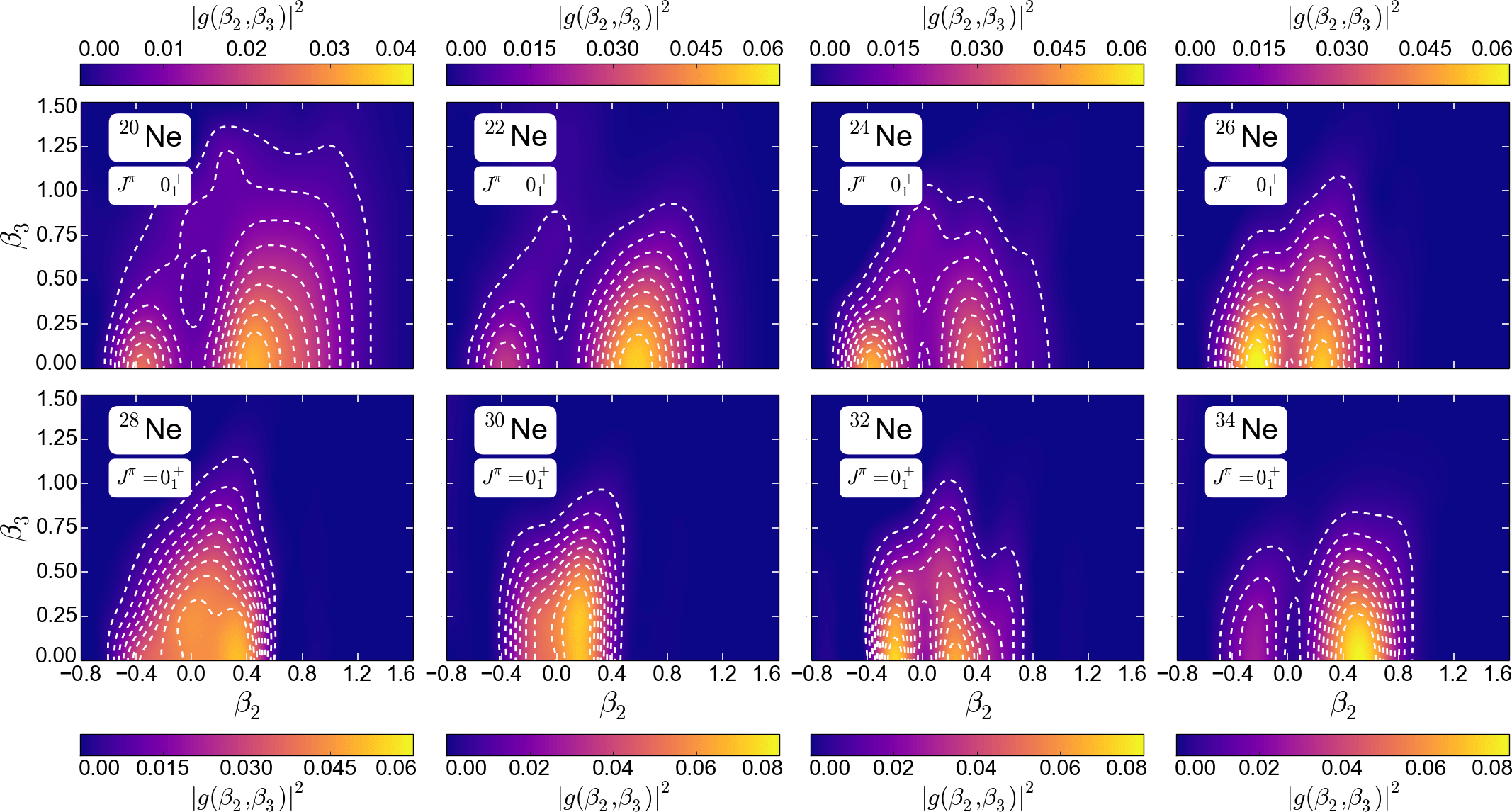}
    }
    \caption{(Color online) Amplitudes of collective wave functions squared $|g(\beta_2, \beta_3)|^2$ of the ground states of $^{20-34}$Ne isotopes. Dashed contours in the $\beta_2 - \beta_3$ plane successively denote a $10\%$ decrease starting from the largest value of the amplitude.}    
    \label{fig:collground}
\end{figure*}

\begin{figure*}[]  
      {\includegraphics[width=0.48\textwidth]{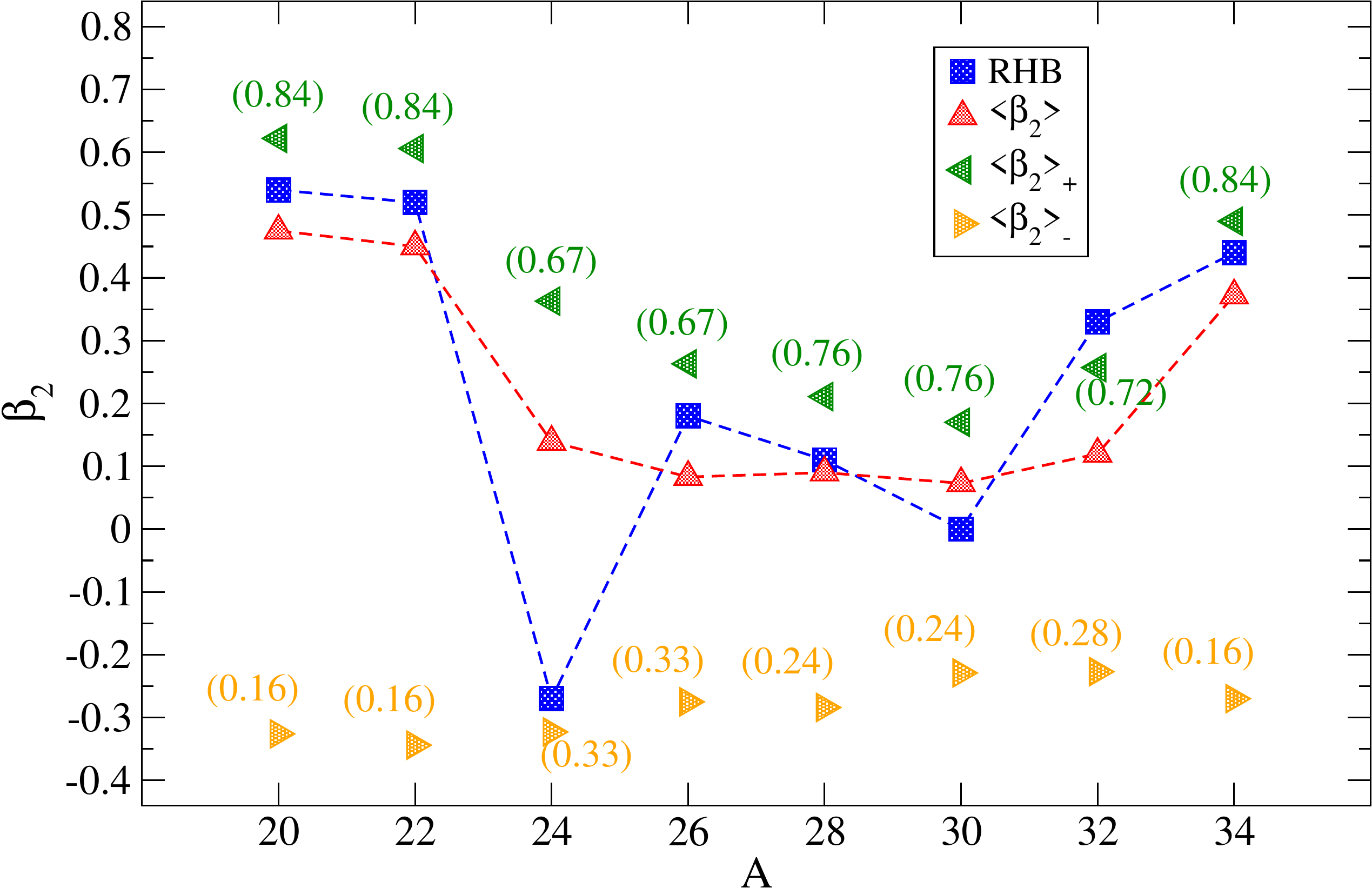}
      \includegraphics[width=0.48\textwidth]{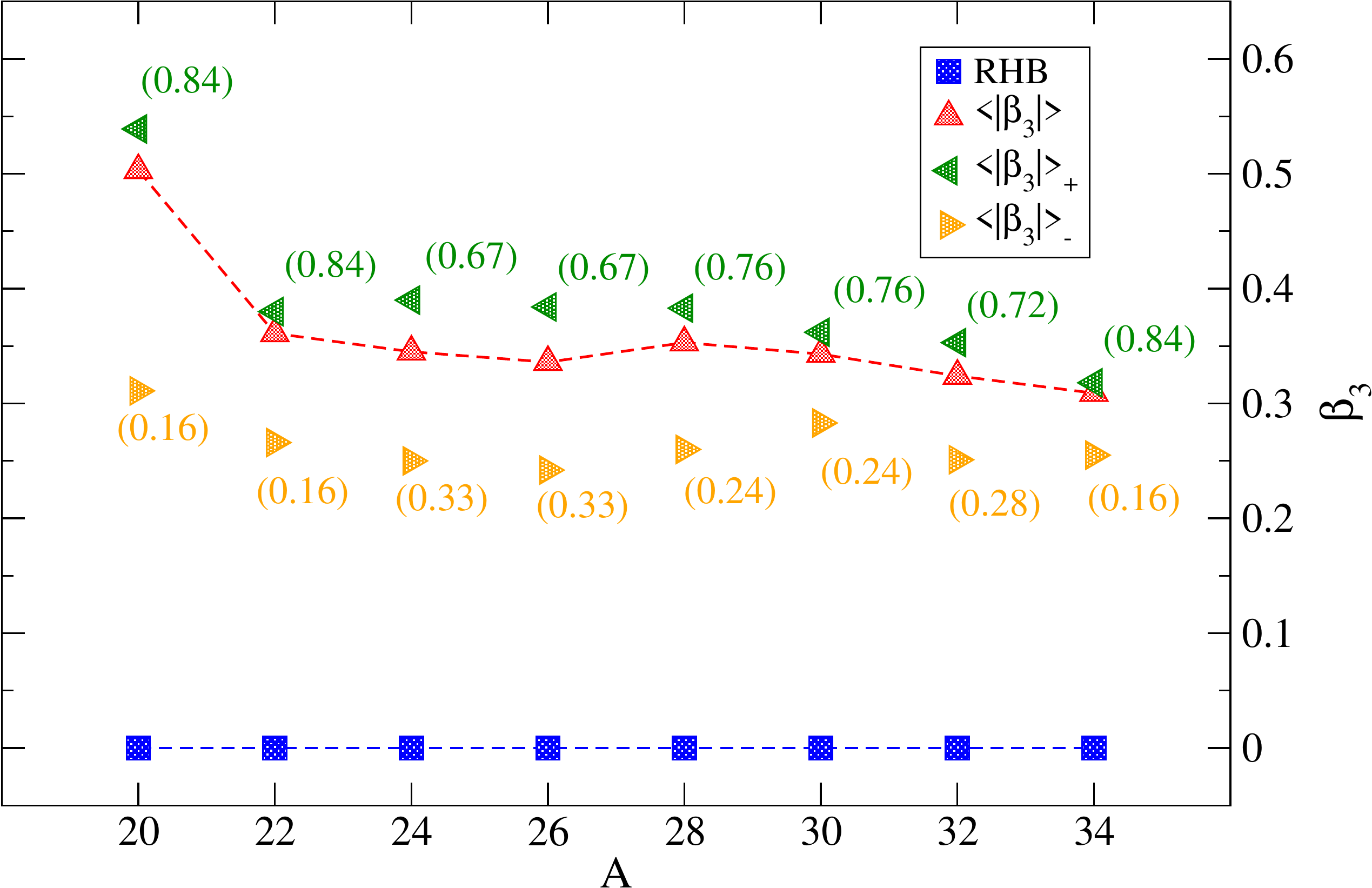}
      }
     \caption{(Color online)Deformation parameters $\beta_2$ (left) and $\beta_3$ (right) that correspond to RHB mean-field minima of $^{20-34}$Ne, in  comparison with the expectation values  $\langle \beta_2 \rangle$ and $\langle |\beta_3| \rangle$ in the corresponding angular momentum- and parity-projected GCM ground states. The deformations obtained by taking the expectation values over only prolate (triangle left) and only oblate (triangle right) configurations, as well as their respective contributions, are also shown.}    
    \label{fig:deformation}
\end{figure*}

\begin{figure*}[]  
      {\includegraphics[width=0.475\textwidth]{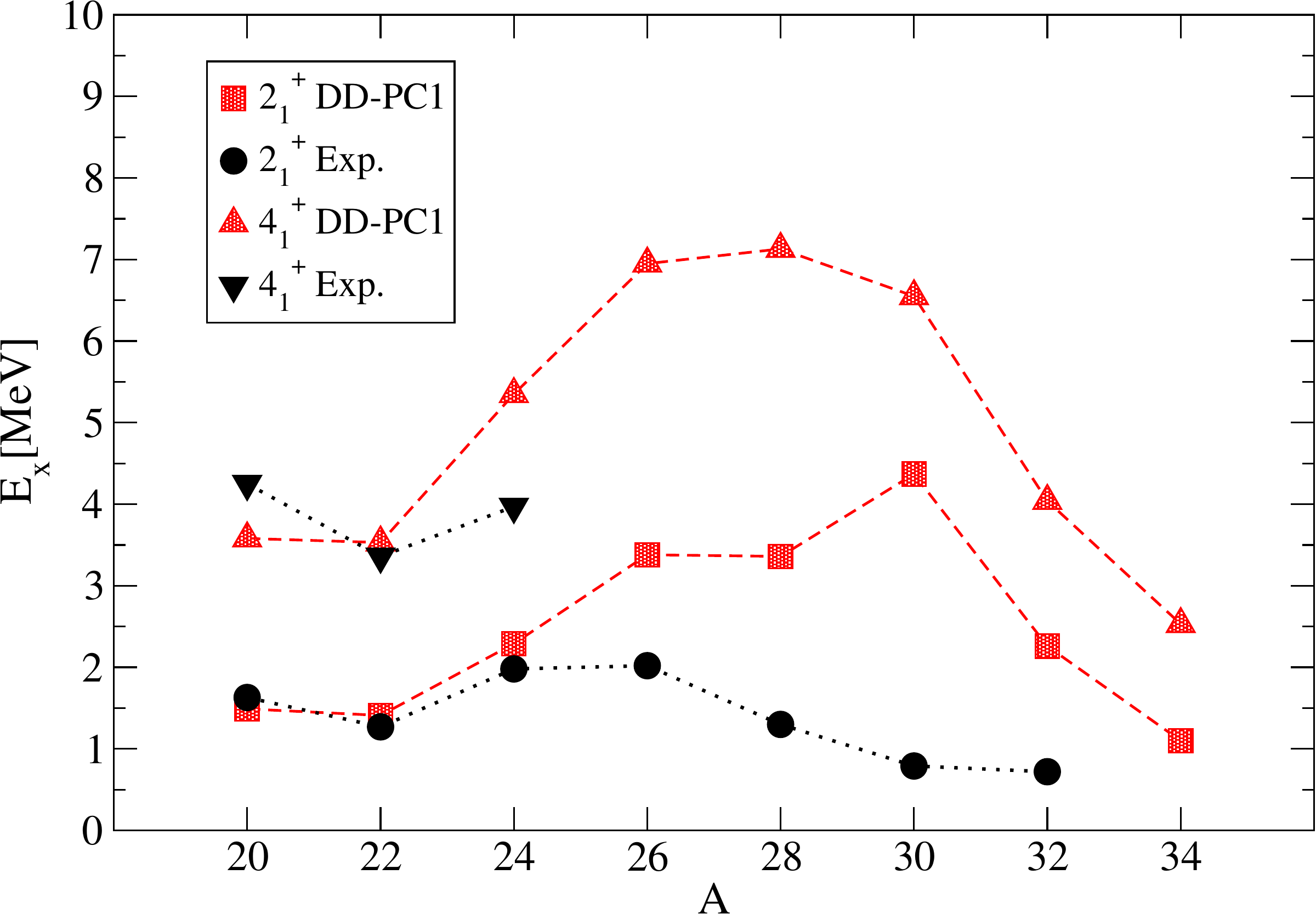}
      \includegraphics[width=0.485\textwidth,height=0.2505\textheight]{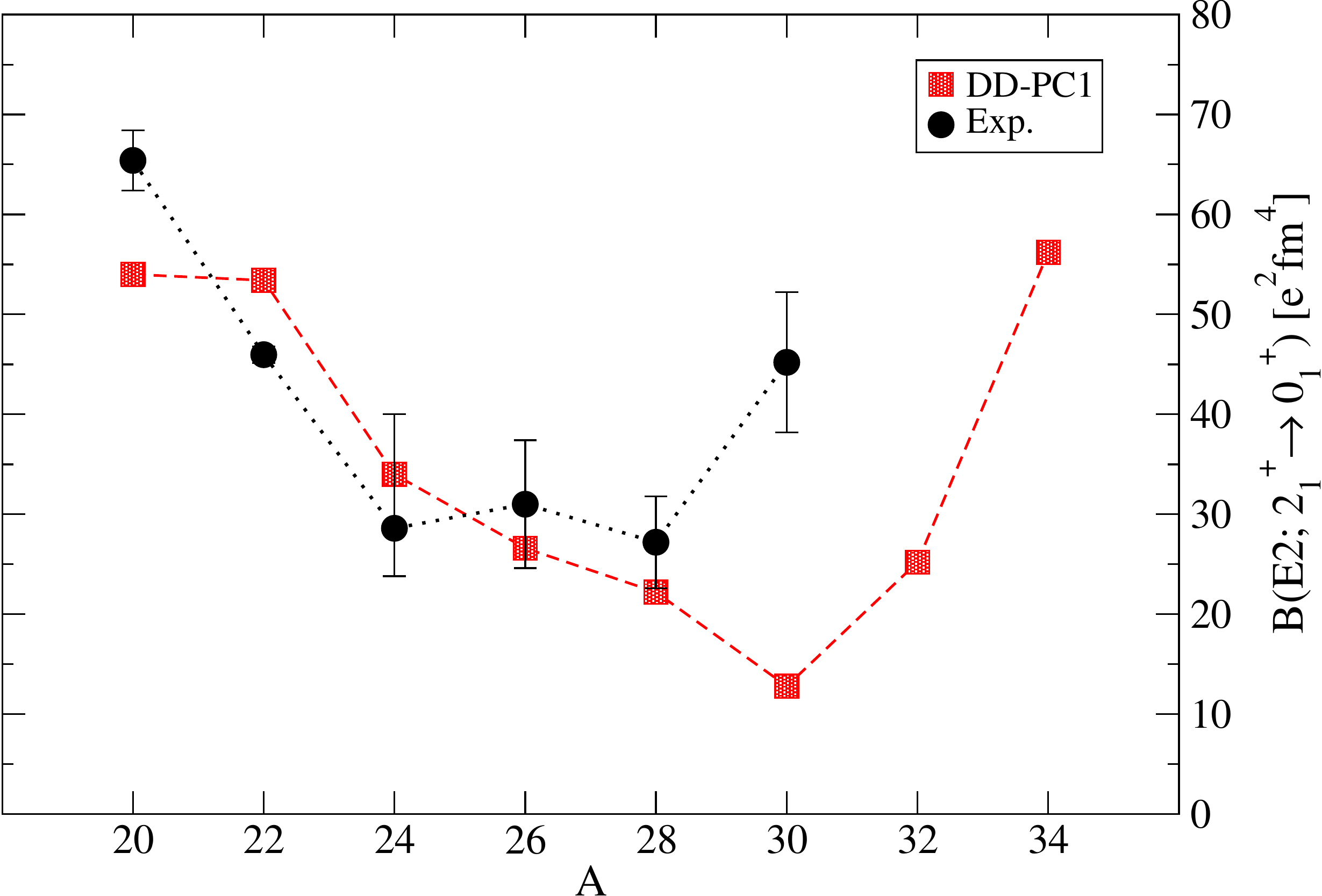}
      }
     \caption{(Color online) Calculated excitation energies of the states $2_1^+$ and $4_1^+$ (left panel) and the transition probabilities $B(E2;2_1^+\rightarrow 0_1^+)$ (right panel) in the even-even $^{20-34}$Ne isotopes, compared with the available experimental data \cite{pritychenko16}.}    
    \label{fig:be2}
\end{figure*}

Even though the ground-state spectroscopic quadrupole moments identically vanish in even-even nuclei, 
it is instructive to calculate the expectation values of the quadrupole deformation parameter
in the correlated ground state:
\begin{equation}
\langle \beta_2 \rangle_{\alpha}^{J \pi} = \sum_i |g_{\alpha}^{J\pi}(q_i)|^2  {\beta_2}_i,
\label{eq:betaaverage}
\end{equation}
where $g_{\alpha}^{J\pi}(q_i)$ denotes the collective wave function (cf. Eq. (\ref{eq:gfun})). In Fig.~\ref{fig:collground} we display the amplitudes squared of the ground-state collective wave functions for $^{20-34}$Ne. This quantity is not an observable, but still it provides useful insight into the structure of correlated ground states. In contrast to the mean-field RHB equilibrium minimum which corresponds to a single configuration in the $(\beta_2,\beta_3)$ plane, the amplitude of the ground state collective wave function manifests the degree of shape fluctuations in both quadrupole and octupole directions. 
In the left panel of Fig.~\ref{fig:deformation} we plot the average $\beta_2$ deformation values 
(\ref{eq:betaaverage}) for $^{20-34}$Ne isotopes in comparison to the deformations that correspond to the self-consistent mean-field RHB minima.
Since the contributions of oblate-deformed configurations to the total collective wave functions are larger than 
$15\%$ over the entire Ne isotopic chain, we additionally display the $\beta_2$ deformations obtained by averaging over only prolate (left triangle) and oblate (right triangle) configurations. In parenthesis we include the respective contributions to the average $\beta_2$ deformation calculated from both prolate and oblate configurations. One notices that oblate configurations give a non-negligible contribution for all isotopes, and this contribution is more pronounced in $^{24-32}$Ne. The nucleus $^{24}$Ne, which exhibits nearly-degenerate oblate and prolate minima on the mean-field level, preserves this structure even after symmetry restoration and configuration mixing. In particular, the dominant component is still prolate-deformed and peaks at 
$\beta_2 \approx 0.4$, but about $1/3$ of the collective wave function spreads over the oblate side and peaks at 
$\beta_2 \approx -0.3$. A similar behaviour is also found in the $^{26}$Ne isotope. The semi-magic nucleus $^{30}$Ne is found to be very weakly prolate-deformed, in contrast to the large ground-state quadrupole deformation deduced from experiment \cite{sorlin08}. By removing two neutrons, the nearly-spherical structure of the ground state appears to be preserved in $^{28}$Ne. The addition of two neutrons, however, leads to the formation of a barrier at the spherical configuration of $^{32}$Ne and a shape-coexisting structure in the collective wave function appears again. In the right panel of Fig.~\ref{fig:deformation} we plot the corresponding  values of the octupole deformation parameter in the RHB minima and correlated ground states, calculated analogously to Eq.~(\ref{eq:betaaverage}). Since $\langle \beta_3 \rangle$ vanishes identically for all collective states with good parity, we plot instead the average values of the corresponding modulus, that is, the $\langle |\beta_3| \rangle$ values. The $\langle |\beta_3| \rangle$ values quantify the role of octupole deformation in the analyzed ground states. Obviously, octupole collectivity is very pronounced in $^{20}$Ne, while it is somewhat weaker and approximately constant over the rest of the isotopic chain, with the average $\langle |\beta_3| \rangle$ value varying between $0.30$ and $0.35$.

\begin{table}[t]
\setlength{\tabcolsep}{4pt}
\renewcommand{\arraystretch}{1.1}
\caption{Calculated ground-state band spectroscopic quadrupole moments (in e fm$^2$) for $J^{\pi} = 2^+, 4^+, 6^+$ in the even-even $^{20-34}$Ne isotopes.}
\centering
\begin{tabular}{cccc}
\hline \hline
$J_{\alpha}^{\pi}$ & $2_1^+$ & $4_1^+$ & $6_1^+$   \\ \hline
$^{20}$Ne & $-16.61$  & $-19.85$  &  $-20.96$ \\
$^{22}$Ne & $-15.01$  & $-18.89$  &  $-20.27$  \\
$^{24}$Ne & $-6.72$   & $-16.42$  &  $-20.55$  \\
$^{26}$Ne & $-9.59$   & $-16.85$  &  $-19.80$ \\
$^{28}$Ne & $-4.43$   & $-17.70$  &  $-24.01$ \\
$^{30}$Ne & $-13.59$  & $-19.72$  &  $-22.25$ \\
$^{32}$Ne & $-13.79$  & $-18.02$  &  $-19.62$ \\
$^{34}$Ne & $-15.86$  & $-20.20$  &  $-21.52$ \\ \hline \hline
\end{tabular}
\label{tab:q2spec}
\end{table} 

In Table~\ref{tab:q2spec} we display the calculated spectroscopic quadrupole moments for the ground-state bands ($J^{\pi} = 2^+, 4^+, 6^+$ states) of $^{20-34}$Ne.  The theoretical values for the $2_1^+$ states in $^{20}$Ne and $^{22}$Ne are in fair agreement with the experimental results: $-23 \pm 3$ efm$^2$ for $^{20}$Ne and $-19 \pm 4$ efm$^2$ for $^{22}$Ne \cite{stone05}. The $2_1^+$ states in $^{24-28}$Ne isotopes are built on either mixed prolate and oblate configurations or weakly-deformed prolate ground states, thus yielding somewhat smaller absolute values for the spectroscopic quadrupole moments. 
Increasing angular momentum stabilizes the prolate-deformed shapes and this is consistent with the larger absolute values for the spectroscopic quadrupole moments of the $4_1^+$ and $6_1^+$ states. 

\begin{figure*}[]  
      {\includegraphics[width=0.495\textwidth]{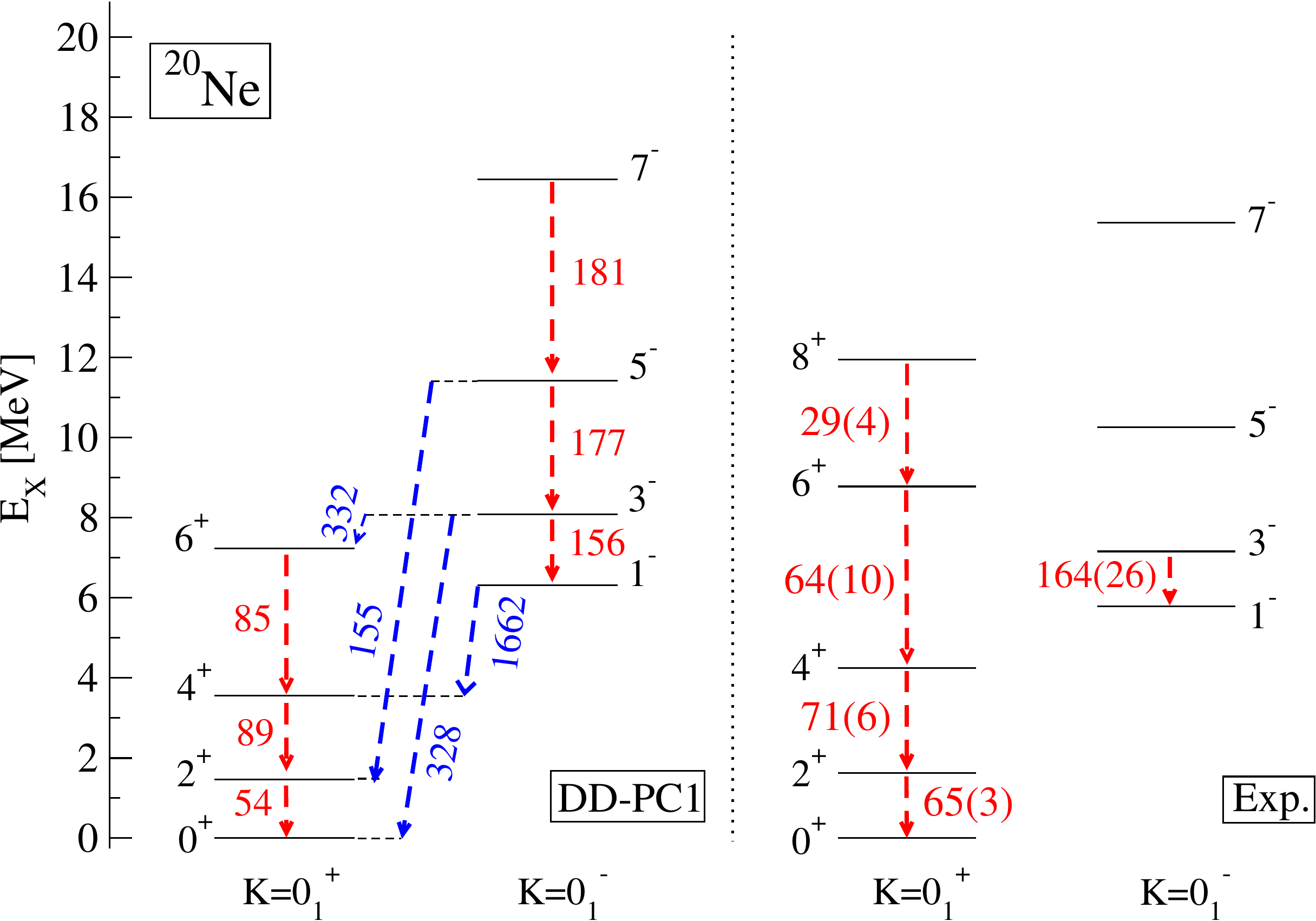}
      \includegraphics[width=0.455\textwidth]{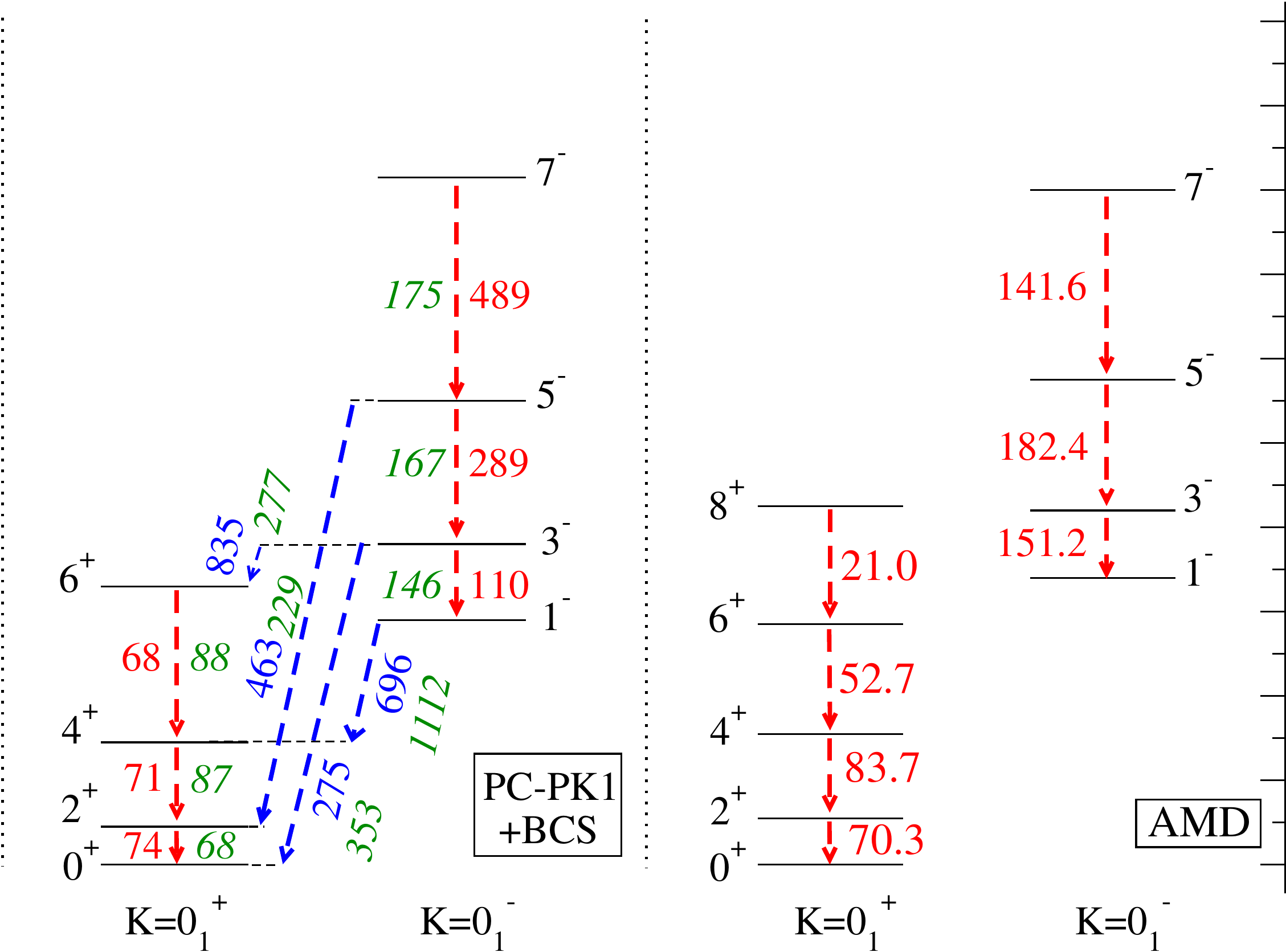}
      }
     \caption{(Color online) Theoretical low-energy excitation spectrum of $^{20}$Ne compared with available data. The calculated $E2$ transition probabilities within the bands (red color, in e$^2$fm$^4$), and $E3$ transition probabilities between the bands (blue color, in e$^2$fm$^6$) are also shown. Results obtained with two other {\em ``beyond mean-field"} models \cite{zhou16,kimura04} are also shown for comparison. See text for details.}    
    \label{fig:ne20spectrum}
\end{figure*}

Finally, in the left panel of Fig.~\ref{fig:be2} we plot the calculated excitation energies for the $2_1^+$ and $4_1^+$ states of $^{20-34}$Ne in comparison to the available experimental values \cite{pritychenko16}. Our results for the lighter isotopes $^{20-24}$Ne are in rather good agreement with data. However, when approaching the $N=20$ neutron shell the theoretical results begin to diverge from experiment, and this is especially pronounced in the $^{30}$Ne isotope. This discrepancy originates from the fact that the functional DD-PC1 predicts the $N=20$ neutron shell closure even for the very neutron-rich isotopes. On the other hand, the breakdown of the $N=20$ neutron magic number, leading to the large quadrupole deformation in the ground state of the $^{30}$Ne isotope, is experimentally a well-established phenomenon~\cite{sorlin08}. We notice that a similar problem occurred in a previous study of $^{32}$Mg based on the relativistic functional PC-F1 \cite{niksic06}, and also in some calculations based on non-relativistic EDFs, e.g. the SLy4 effective interaction~\cite{heenen00}. In addition, the present study is restricted to axial shapes, whereas in some of the heavier Ne isotopes additional degrees of freedom, e.g. triaxial, could be important. Similar results were also obtained in the study of quadrupole collectivity of neutron-rich Neon isotopes based on the Gogny force~\cite{rodriguez03}.
The calculated  $B(E2; 2_1^+ \rightarrow 0_1^+)$ transition probabilities in the Ne isotopic chain are displayed in the right panel of
Fig.~\ref{fig:be2}, and compared with the available data~\cite{pritychenko16}. The theoretical results reproduce the experimental values except for $^{30}$Ne. Because of the predicted $N=20$ neutron shell closure, the calculated $B(E2; 2_1^+ \rightarrow 0_1^+)$ value is much smaller than the corresponding experimental value. 

\subsection{The self-conjugate nucleus $^{20}$Ne}

\begin{figure*}[]  
      {\includegraphics[scale=0.32]{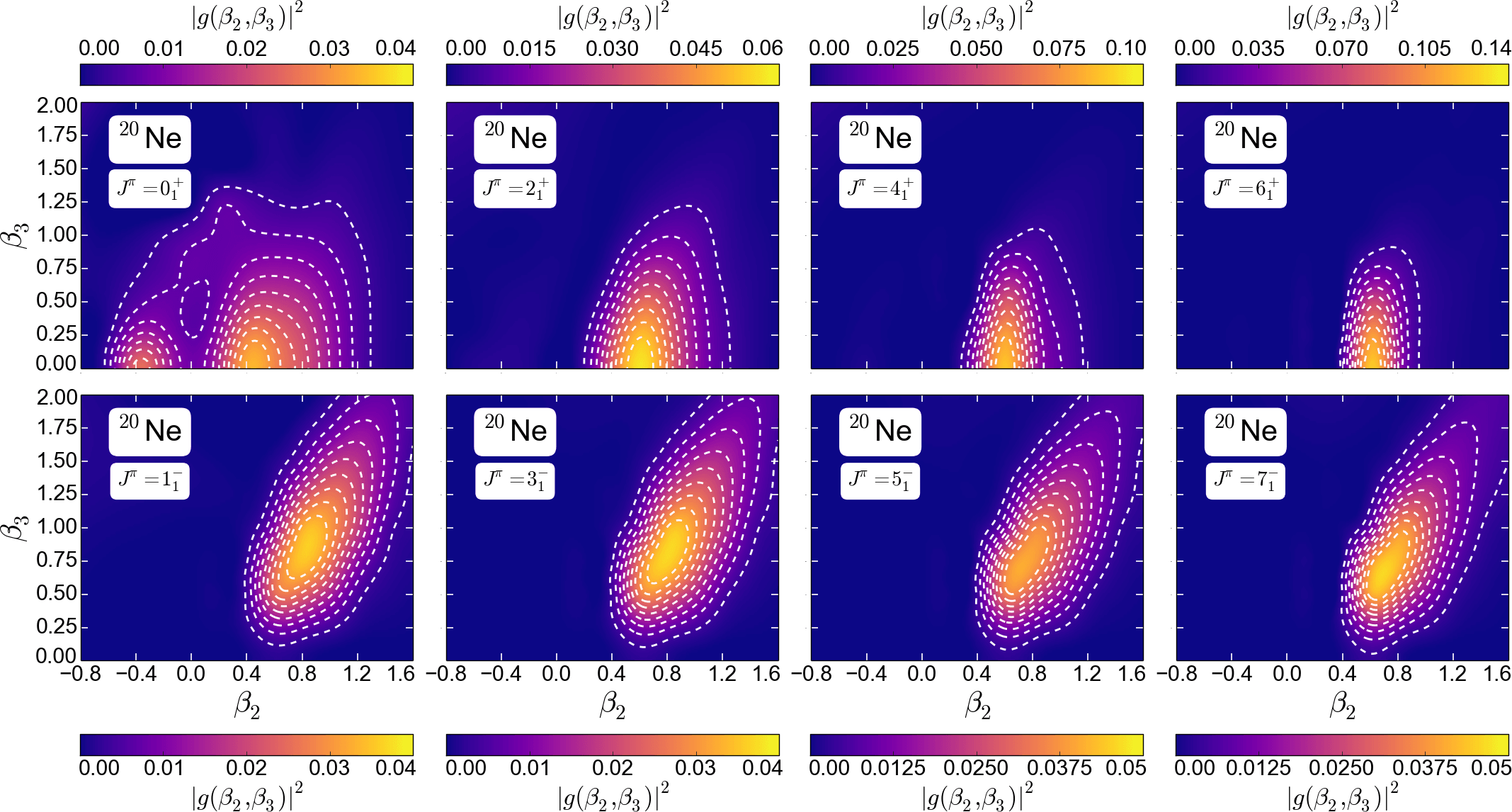}
          }
    \caption{(Color online) Amplitudes of collective wave functions squared $|g(\beta_2, \beta_3)|^2$ of the low-energy levels of $^{20}$Ne. Dashed contours 
    in the $\beta_2 - \beta_3$ plane successively denote a $10\%$ decrease starting from the largest value of the amplitude.}    
    \label{fig:collNe20}
\end{figure*}

$^{20}$Ne presents a very interesting example of a nucleus that exhibits admixtures of cluster configurations already in the ground state. Previous studies based on the relativistic EDF framework have shown that the reflection-asymmetric $\alpha+^{16}$O structure indeed appears already on the mean-field level~\cite{ebran14a,ebran14b}. However, to obtain a quantitative description of the low-energy structure of $^{20}$Ne, correlations related to symmetry restoration and shape fluctuations have to be taken into account. 
In Fig.~\ref{fig:ne20spectrum} we display the calculated low-lying spectrum of $^{20}$Ne in comparison to available data, and predictions of two other theoretical {\em ``beyond mean-field"} studies. The results of the present calculation are shown in the first column, and the experimental excitation spectrum in the second. The third column includes results obtained in a recent study of $^{20}$Ne based on the relativistic EDF PC-PK1~\cite{zhou16}. In contrast to the present analysis, in Ref.~\cite{zhou16} pairing correlations were treated in the BCS approximation, and configuration mixing calculation was performed using a set of $54$ prolate-deformed mean-field basis states with projection on angular momentum, parity and particle number. 
In addition to this set of basis states, denoted as \emph{full configuration}, an additional set was considered that contains only $6$ configurations whose mixing yields optimal results in comparison to experiment. This set of basis states was denoted as \emph{optimal configuration} and contains four prolate configurations, one oblate configuration, and the spherical configuration. Although the excitation energies obtained with both sets of basis states are very similar, one finds significant differences in the calculated transition probabilities. Results obtained with the \emph{optimal configuration} set are shown in green in Fig.~\ref{fig:ne20spectrum}, whereas those obtained with the \emph{full configuration} set are shown in red (intra-band) and blue (inter-band). Note that the present GCM calculation uses a total of $157$ configurations, both oblate and prolate. Finally, in the fourth column we show results obtained using the deformed basis antisymmetrized molecular dynamics model (AMD)~\cite{kimura04}. This model employs a triaxially-deformed Gaussian function for the spatial part of the single-particle wave packet and, although the formation of cluster states is not assumed {\em a priori} in this model, nucleon localization is inbuilt by using Gaussian wave packets. 
 
The yrast-band energies are reproduced reasonably well by all three models, with the excitation spectra somewhat compressed in comparison to the experiment. The two GCM models slightly underestimate the moment of inertia for the negative-parity band, that is, the energy levels in the negative-parity band are a bit spread out compared to the experimental values. The best agreement with data for the transition rates within the yrast band is obtained with the AMD model. Even though the present calculation predicts a marginally smaller $B(E2; 2_1^+ \rightarrow 0_1^+)$ value, and overestimates the  $B(E2; 4_1^+ \rightarrow 2_1^+)$ and $B(E2; 6_1^+ \rightarrow 4_1^+)$ values, it reproduces the overall trend for the $B(E2)$ values within the yrast band. On the other hand, the PC-PK1 calculation with the 
\emph{full configuration} set fails to reproduce the observed increase in the $E2$ transition probabilities from 
$2_1^+ \to 0_1^+$ to $4_1^+ \to 2_1^+$, while the same model with the \emph{optimal configuration} set does not reproduce the decrease in the $E2$ transition probabilities from $4_1^+ \to 2_1^+$ to $6_1^+ \to 4_1^+$ transitions.

\begin{figure*}[]  
      {\includegraphics[width=0.8\textwidth]{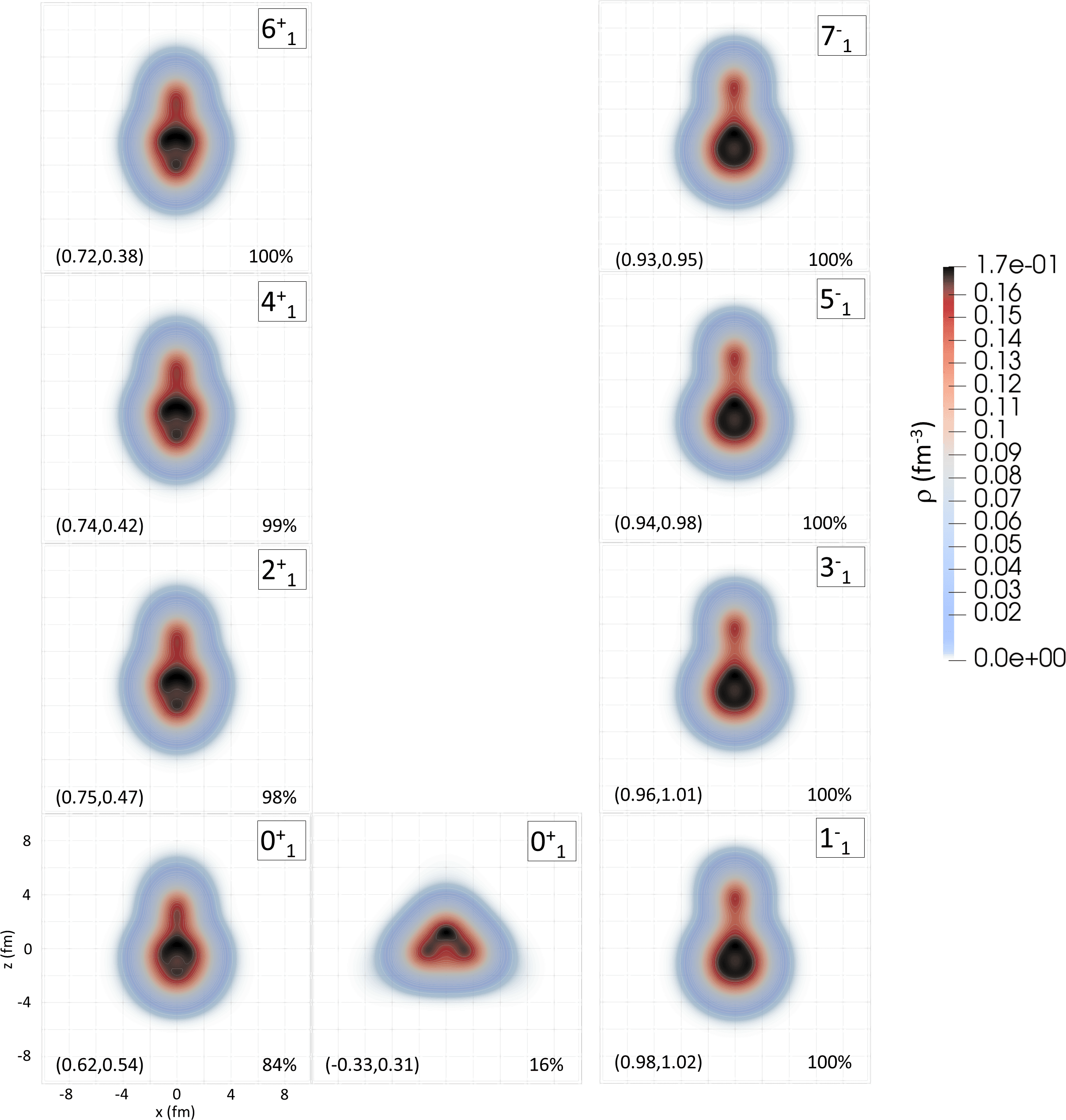}
      }
     \caption{(Color online) Characteristic intrinsic nucleon densities of states of the ground-state band and the $K^{\pi}=0^-$ band in $^{20}$Ne. See text for description.}    
    \label{fig:ne20dens}
\end{figure*}

These differences between the present calculation and the PC-PK1 - based study could probably be attributed to the different selection of basis states used in the configuration mixing calculation. In the upper row of Fig.~\ref{fig:collNe20} we plot the amplitudes of collective wave functions squared for the yrast-band states in the  $(\beta_2,\beta_3)$ plane. One notices that, while the collective wave function for $0_1^+$ displays a significant contribution from oblate configurations, the wave functions of the states with higher angular momenta are concentrated around the prolate deformation $\beta_2 \approx 0.5$. Omitting oblate configurations from the basis space (\emph{full configuration} set in the PC-PK1 calculation) will produce a prolate-deformed ground state hence overestimating the $B(E2; 2_1^+ \rightarrow 0_1^+)$ value. By including just one oblate configuration (\emph{optimal configuration} set in the PC-PK1 calculation) this value is reduced.

The $E2$ transition probabilities for  the $K^\pi=0^-$ band obtained in the present study agree with the AMD calculation, particularly for the $3_1^- \to 1_1^-$ and $5_1^- \to 3_1^-$ transitions. One also notices the agreement between the predicted and experimental $B(E2; 3_1^- \rightarrow 1_1^-)$ value. On the other hand, the transition probabilities obtained in the PC-PK1 calculation based on the \emph{full configuration} set differ considerably from the other two models and experiment. This problem can be resolved by including one oblate configuration (\emph{optimal configuration} set in the PC-PK1 calculation) in the basis set. In the lower row of Fig.~\ref{fig:collNe20} we plot the amplitudes of collective wave functions squared for the
negative-parity band, and these can be compared with the right column in Fig. 5 of Ref.~\cite{zhou16}, where the same amplitudes were calculated using the PC-PK1 interaction with the \emph{full configuration} set. We notice that the present calculation predicts that all wave functions are concentrated around $(\beta_2 \approx 0.8, \beta_3 \approx 0.7)$, while the study of 
Ref.~\cite{zhou16} predicts a broader distribution of the wave functions of the $K^\pi = 0^-$ band, with the peak position shifting towards smaller values of $\beta_2$ with increasing angular momentum. 
Finally, our predictions for the $E3$ transition probabilities between the $K^\pi=0^-$ and $K^\pi=0^+$ bands are in fair agreement with the results obtained in Ref.~\cite{zhou16} using the  \emph{optimal configuration} set.

\begin{figure*}[]
   {\includegraphics[width=0.49\textwidth]{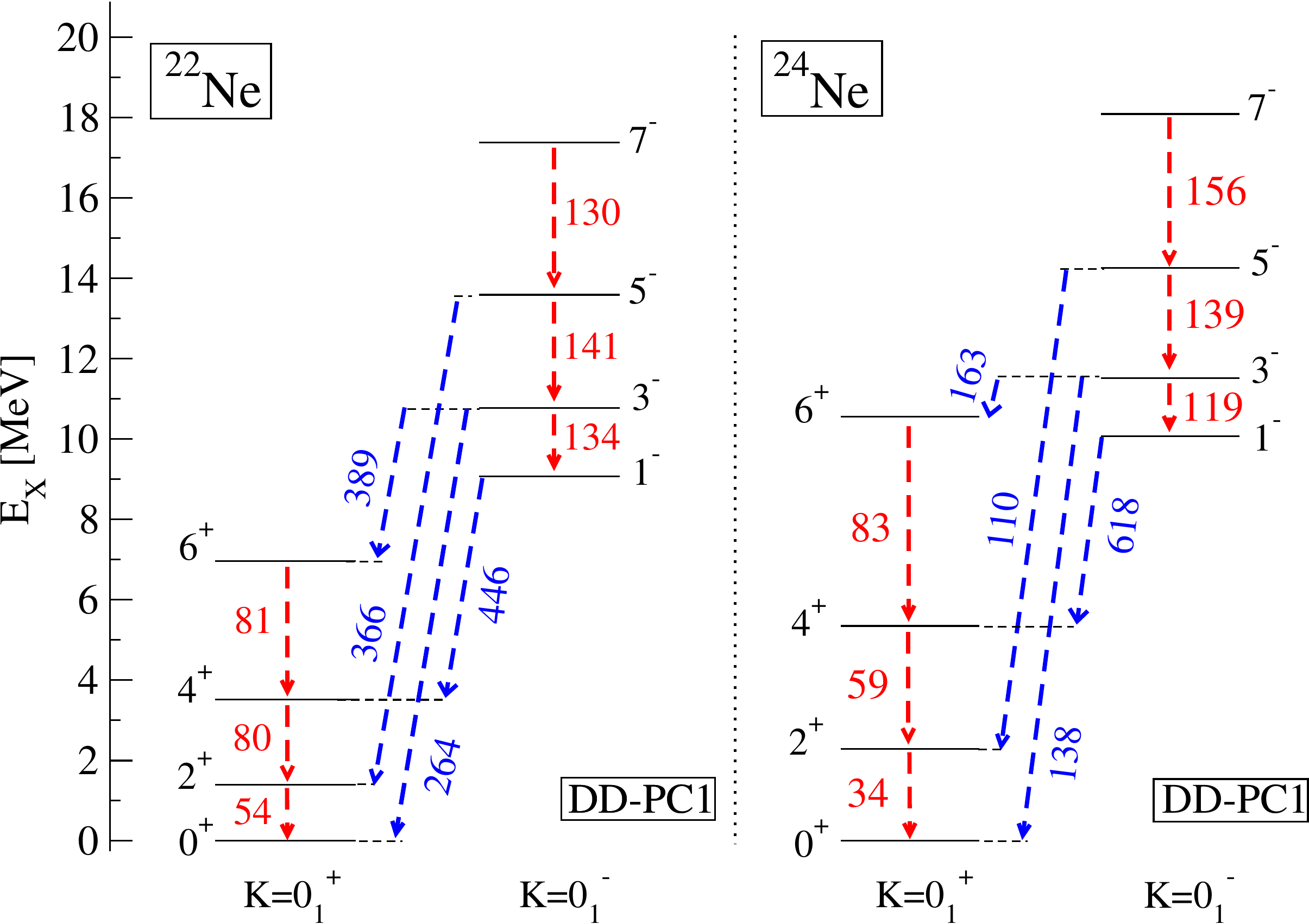}
      \includegraphics[width=0.455\textwidth]{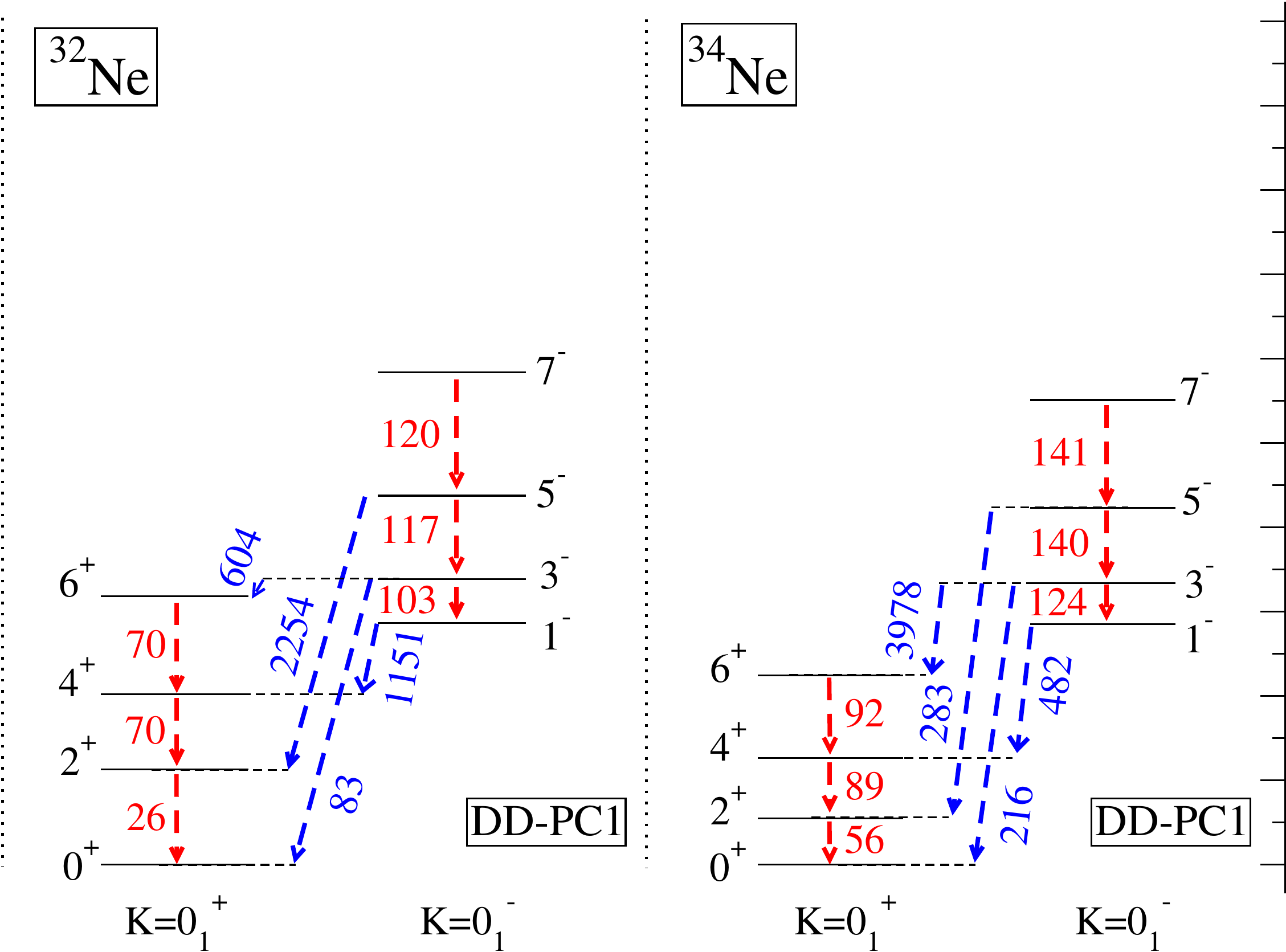}
      }
     \caption{(Color online) Calculated low-energy spectra of $^{22,24}$Ne (left panel (a)) and $^{32,34}$Ne (right panel (b)). The $E2$ reduced transition probabilities within the bands (red color, in e$^2$fm$^4$) and $E3$ transition probabilities between the bands (blue color, in e$^2$fm$^6$) are also shown.}    
    \label{fig:nespectra}
\end{figure*}

To illustrate the evolution of cluster structures in the collective states, in Fig. \ref{fig:ne20dens} we display the characteristic intrinsic nucleon densities of the ground-state band and the $K^{\pi}=0^-$ band in $^{20}$Ne. For each state, the corresponding prolate and/or oblate deformation parameters $(\beta_2, \beta_3)$, shown in parenthesis, are calculated by averaging over the prolate-deformed and oblate-deformed configurations separately (see eq.~(\ref{eq:betaaverage})). For the 
average prolate or oblate $(\beta_2, \beta_3)$ we plot the corresponding intrinsic total nucleon density in the {\em xz} plane. These densities are obtained by 
axial RHB calculations constrained to the average $(\beta_2, \beta_3)$. In each panel we also include the percentage of prolate or oblate configurations in the 
collective wave function. Only in the ground state there is a significant contribution of oblate configurations, while for all other yrast states the intrinsic structure is dominated by prolate configurations. The major contribution to the $^{20}$Ne ground state comes from the reflection-asymmetric prolate-deformed $\alpha+^{16}$O configuration, but it also contains a 16\% admixture of oblate-deformed configurations with a characteristic intrinsic density resembling the $2\alpha+^{12}$C structure. 
The predicted transitional character of $^{20}$Ne ground state between mean-field and cluster-like structures is in agreement with AMD analyses \cite{beck12,kimura04}. It is remarkable that, starting from a basis of more than $150$ mean-field states, the GCM calculation brings out the two main components of the collective state that are used as \emph{a priori} basis states in custom built cluster models. The transitional nature of the ground state is usually invoked to explain the relatively high excitation energy of its parity-doublet $1_1^-$ state, which is predicted to exhibit a pronounced $\alpha+^{16}$O structure by both the present study and AMD calculations. Increasing angular momentum leads to a gradual dissipation of the $\alpha+^{16}$O structure in the $K^{\pi}=0^-$ band. However, this process appears to develop faster in the AMD \cite{kimura04} and PC-PK1+BCS \cite{zhou16} calculations. A similar trend is observed in the ground-state band, particularly for the $J^{\pi}=6^+$ state shown in Fig.~\ref{fig:ne20dens}, for which a weak $\alpha-^{12}$C$-\alpha$-like structure appears.


\subsection{Neutron-rich Ne isotopes}

In the remainder of this study we analyze the structure of a selected set of heavier Ne isotopes: $^{22,24}$Ne and $^{32,34}$Ne. In the left panel of Fig.~\ref{fig:nespectra} the excitation spectra of $^{22}$Ne and $^{24}$Ne are shown. Compared to the spectrum of $^{20}$Ne, adding two neutron does not significantly modify the collective structure and both the energy spectrum and the transition rates in $^{22}$Ne are very similar to the ones predicted for $^{20}$Ne. The yrast band of $^{24}$Ne, however, is considerably more stretched compared to both $^{20}$Ne and $^{22}$Ne. Shape mixing in the ground-state band of the $^{24}$Ne isotope is also reflected in the reduced transition probabilities. The lowest negative-parity bands, on the other hand, are rather similar in all three isotopes $^{20,22,24}$Ne. The $K^{\pi}=0_1^-$ excitation energies increase with neutron number, and the $E2$ transition rates generally decrease, reflecting a reduction of octupole collectivity.

\begin{figure*}[]
 \begin{center} 
\includegraphics[width=0.9\textwidth,height=0.4\textheight]{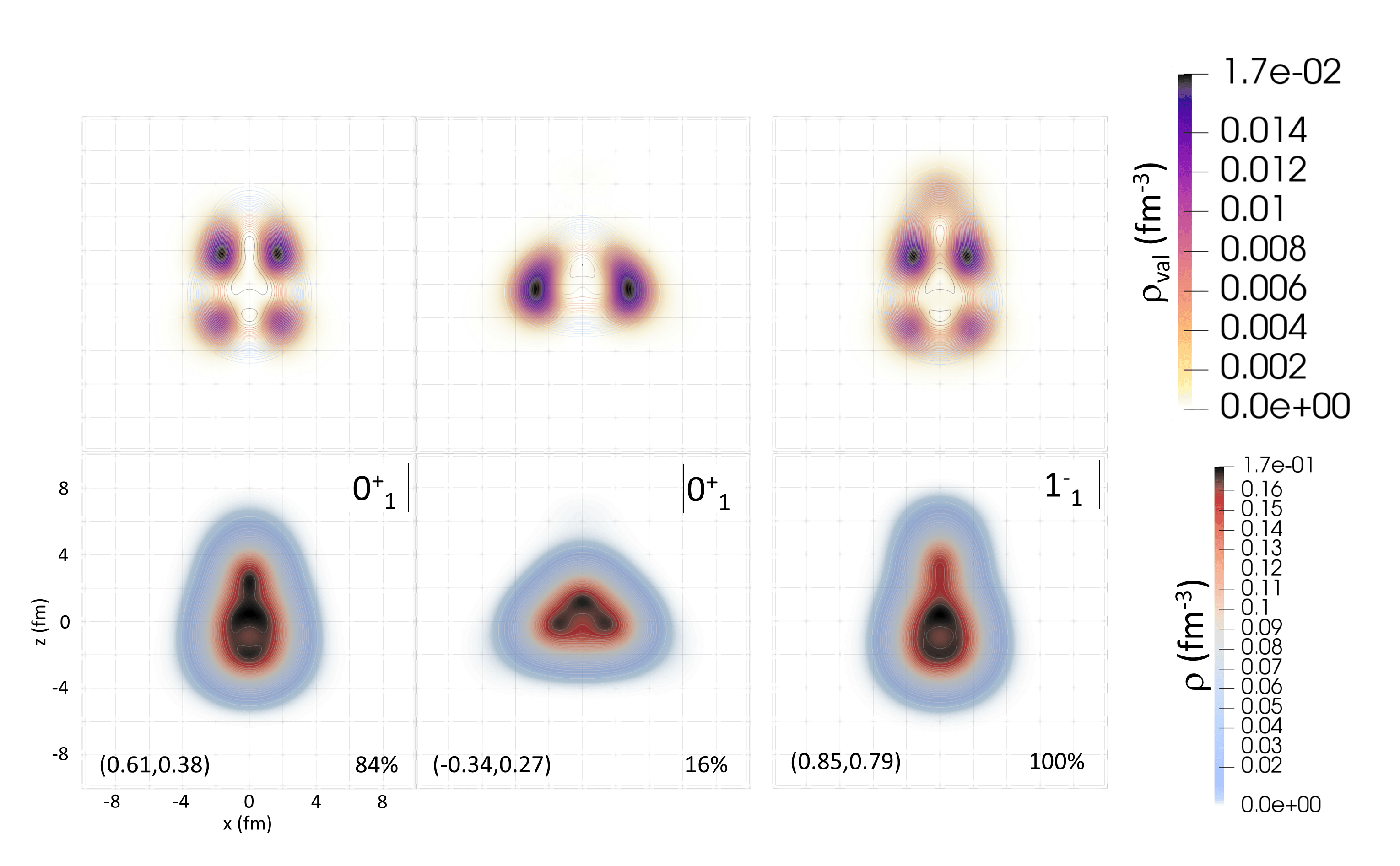}
     \caption{(Color online) Characteristic intrinsic nucleon densities of the ground state and the $K^{\pi}=0_1^-$ band-head in $^{22}$Ne. Total nucleon densities (lower panel) and valence neutrons densities (upper panel) are shown.}    
    \label{fig:ne22dens} 
    \end{center} 
\end{figure*}

\begin{figure*}[]
 \begin{center} 
  \includegraphics[width=0.92\textwidth,height=0.4\textheight]{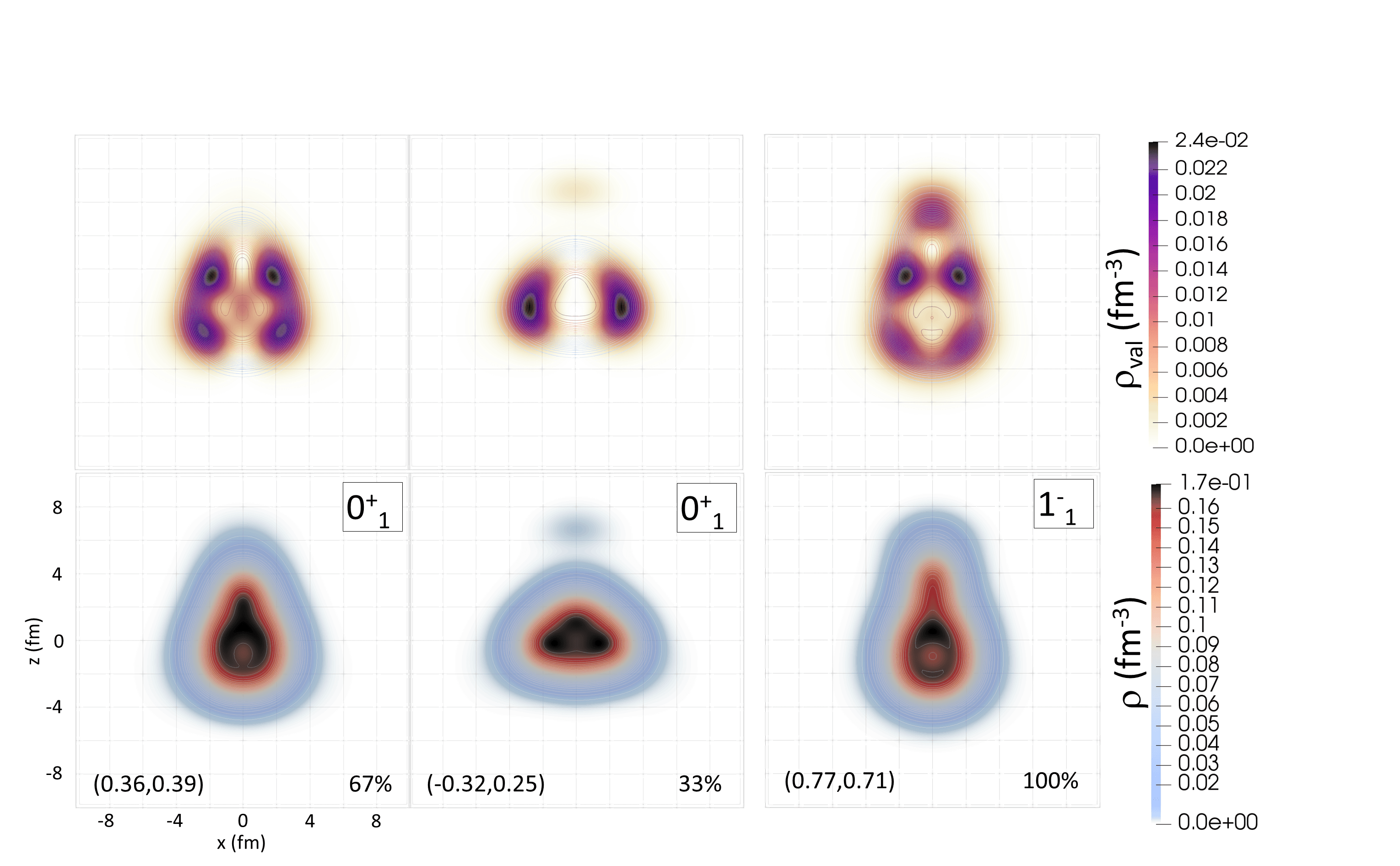} 
   \caption{(Color online) Same as in the caption to Fig.~\ref{fig:ne22dens}, but for $^{24}$Ne.}    
    \label{fig:ne24dens}
     \end{center} 
\end{figure*}

Further insight into the structure of the lowest positive- and negative-parity collective states is gained from the characteristic intrinsic nucleon densities, determined as described in the previous subsection. 
In the lower panels of Figs.~\ref{fig:ne22dens} and \ref{fig:ne24dens} we plot the total intrinsic nucleon densities that correspond to the average prolate and oblate deformations of the ground state, and average prolate deformation of the $K^{\pi}=0_1^-$ band-head in $^{22}$Ne and $^{24}$Ne isotopes, respectively. The entire ground-state band of $^{22}$Ne exhibits a structure very similar to the one of $^{20}$Ne, with slightly reduced values of the octupole deformation. 
For $^{24}$Ne, in addition to the ground state, shape coexistence is rather pronounced also in the $2_1^+$ and $4_1^+$ states. In particular, oblate-deformed configurations account for $35\%$ and $12\%$, respectively, of the corresponding collective wave functions. It is also interesting to point out that, even though the prominent $\alpha+^{16}$O structure is predicted in the negative-parity bands of both $^{22}$Ne and $^{24}$Ne, for the latter the calculated octupole deformation does not decrease for higher angular momenta and the quadrupole deformation in fact increases, that is, the opposite trend as compared to $^{20}$Ne and $^{22}$Ne. 

$^{22}$Ne and $^{24}$Ne are also interesting in the context of excess neutrons playing the role of molecular bonding between cluster structures. This analysis is based on the picture of nuclear molecular states, that emerges if the total nucleon density is decomposed into the density of clusters and the density of additional valence neutrons. For covalent bonding, a negative-parity orbital perpendicular to the axis connecting the two clusters is called a $\pi$-orbital, while a positive-parity orbital parallel to this axis is called a $\sigma$-orbital (cf. Fig. 7 of Ref.~\cite{kanada12}). To qualitatively determine the density of the valence neutrons, after solving the relativistic Hartree-Bogoliubov equations for a given deformation $(\beta_2,\beta_3)$, the solution is transformed into the canonical basis which diagonalizes the density matrix~\cite{RS.80}. For the Neon isotopes considered in this analysis, the five deepest proton and neutron orbitals exhibit occupation numbers $n_i > 0.99$, hence their contribution is interpreted as the $^{20}$Ne core density, while the remaining orbitals comprise the valence density. 
In the upper panels of Figs.~\ref{fig:ne22dens} and \ref{fig:ne24dens} we plot the  intrinsic valence neutrons densities that correspond to the average oblate and/or prolate deformations of the ground state and the $K^{\pi}=0_1^-$ band-head in $^{22}$Ne and $^{24}$Ne isotopes. The ground-state band of $^{22}$Ne exhibits a characteristic $\pi$-bonding, in agreement with the AMD analysis of Ref. \cite{kimura07}. The same study predicted a pronounced $\sigma$-bond already for the band-head of the $K^{\pi}=0_1^-$ band. In our case, however, the $1_1^-$ state still appear to preserve the $\pi$-bond-like molecular bonding, and only the increase of angular momentum leads to a development of the $\sigma$-bond, particularly pronounced for the $7_1^-$ state. The situation is different in the $^{24}$Ne isotope, where the ground-state band is characterized by $\pi$-bonding, while the entire negative-parity band exhibits a pronounced $\sigma$-bond.

We conclude the present analysis by focusing on the two most neutron-abundant Neon isotopes, that is, $^{32}$Ne and $^{34}$Ne. Both isotopes are predicted to be stable against two-neutron emission and, moreover, the calculated two-neutron separation energies are found within their respective experimental error bars (see Fig. \ref{fig:senergy}). In the right panel of Fig.~\ref{fig:nespectra} we plot the corresponding low-energy spectra and electromagnetic transition probabilities. The ground-state band spectrum of $^{32}$Ne is rather similar to those of lighter  isotopes, namely $^{20}$Ne and $^{22}$Ne. However, because of a significant presence of oblate deformation in the corresponding ground state, the calculated $E2$ transition probability is rather small and closer to that of the shape-coexisting $^{24}$Ne. The ground-state band spectrum of $^{34}$Ne, which is built on a  strongly prolate-deformed $0^+$ state, is much more compressed. In particular, the energies of the $2^+$ and $4^+$ states are found to be the lowest in the whole isotopic chain, while the corresponding $E2$ transition rate to the ground state is the largest. The negative-parity spectra of these isotopes are rather similar and the corresponding band-heads are found at relatively low energies, indicating pronounced collectivity. Because of the different structure of their ground states, the octupole transition to the ground state in $^{34}$Ne is almost three times larger than the one in $^{32}$Ne. In fact, it is only the ground states of these isotopes that exhibit pronounced prolate-oblate shape coexistence. Excited states, on the other hand, are built on stable prolate and reflection-asymmetric configurations.   

\section{Summary}
\label{sec:conclusion}

The structure of the lowest positive- and negative-parity bands of $^{20}$Ne and the neutron-rich Neon isotopes has been analyzed using a {\em ``beyond mean-field"} approach based on relativistic energy density functionals. Starting from self-consistent axially-symmetric quadrupole and octupole deformed relativistic Hartree-Bogoliubov states, projections on angular momentum and parity are carried out, and the resulting symmetry-conserving states are subsequently used in a configuration mixing calculation that employs the generator coordinate method. This model enables a consistent, parameter-free calculation of excitation spectra and electric transition probabilities, both for the ground-state band as well as for the excited $K^{\pi}=0^{\pm}$ bands. A good agreement with experimental results for the energies of the lowest positive-parity states and for the quadrupole transition rates has been obtained over the chain of isotopes considered, as well as with available data on low-energy negative-parity states. In particular, the spectroscopic properties of $^{20}$Ne have been calculated at a level of accuracy comparable to those obtained using more specific models, such as antisymmetrized molecular dynamics. In addition, the contribution of cluster configurations to the intrinsic nucleon density distributions has been examined and, particularly, the ground state of $^{20}$Ne has been shown to exhibit a transitional character between homogeneous matter and a cluster phase. Furthermore, the low-lying spectra of $^{22}$Ne and $^{24}$Ne have been calculated and their characteristic intrinsic nucleon densities and valence-neutron bonds analyzed, as well as spectroscopic properties of the two isotopes at the neutron drip-line: $^{32}$Ne and $^{34}$Ne. 

The model that has been used in this study, the angular momentum- and parity-projected generator coordinate method, is based on the universal framework of energy density functionals. Rather than using specific effective interactions adjusted to a particular mass region and optimized basis states, it implements functionals that are applicable across the entire nuclear chart, and does not make any assumption about single-nucleon localization. The advantages of using EDFs: global effective interactions, an intuitive interpretation of results in terms of intrinsic shapes, calculations performed in the full space of occupied single-nucleon states, are obvious already at the mean-field level. It is, however, the development of beyond mean-field methods, including collective correlations related to symmetry restoration and shape fluctuations, that enables an accurate description of spectroscopic properties. Of course, global effective interactions might not describe particular properties determined by shell evolution in a specific mass region such as, in this case, the erosion of the $N=20$ shell closure in very neutron-rich nuclei. However, this framework, especially when extended to restore further symmetries, e.g. particle number, and include additional collective variables, presents one of the most promising theoretical tools for studies of the coexistence of the quantum-liquid and cluster states in nuclei.
\acknowledgements{
This work was supported in part by 
the Croatian Science Foundation -- project "Structure and Dynamics
of Exotic Femtosystems" (IP-2014-09-9159) and the QuantiXLie Centre of Excellence,
a project cofinanced by the Croatian Government and European Union through the 
European Regional Development Fund - the Competitiveness and Cohesion Operational
Programme (KK.01.1.1.01).
}

\end{document}